\documentclass[gmd]{copernicus}

\bartext{Model experiment description paper}
\begin{document}
\nolinenumbers
\title{A modern-day Mars climate in the Met Office Unified Model:\hack{\break} dry simulations}

\Author[1]{Danny}{McCulloch}
\Author[1]{Denis~E.}{Sergeev}
\Author[1]{Nathan}{Mayne}
\Author[1]{Matthew}{Bate}
\Author[2]{James}{Manners}
\Author[2,1]{Ian}{Boutle}
\Author[2]{Benjamin}{Drummond}
\Author[1]{Kristzian}{Kohary}

\affil[1]{Department of Physics and Astronomy, University of Exeter, Exeter, EX4 4QL, UK}
\affil[2]{Met Office, FitzRoy Road, Exeter, EX1 3PB, UK}

\correspondence{Danny McCulloch (dm575@exeter.ac.uk)}

\runningtitle{Modern Mars in the Unified Model}

\runningauthor{D. McCulloch et al.}

\received{29 July 2022}
\pubdiscuss{11 August 2022}
\revised{21 December 2022}
\accepted{23 December 2022}
\published{}

\firstpage{1}
\maketitle

\begin{abstract}
We present results from the Met Office Unified Model (UM), a world-leading climate and weather model, adapted to simulate a dry Martian climate. We detail the adaptation of the basic parameterisations and analyse results from two simulations, one with radiatively active mineral dust and one with radiatively inactive dust. These simulations demonstrate how the radiative effects of dust act to accelerate the winds and create a mid-altitude isothermal layer during the dusty season. We validate our model through comparison with an established Mars model, the Laboratoire de Météorologie Dynamique planetary climate model (PCM), finding good agreement in the seasonal wind and temperature profiles but with discrepancies in the predicted dust mass mixing ratio and conditions at the poles. This study validates the use of the UM for a Martian atmosphere, highlights how the adaptation of an Earth general circulation model (GCM) can be beneficial for existing Mars GCMs and provides insight into the next steps in our development of a new Mars climate model.
\end{abstract}

\introduction

Understanding Mars' climate has been the motivation for many missions and numerical models for decades, and through these, many of the mechanisms driving the Martian climate have been unveiled. With our expanding comprehension of Mars' climate, studies have been able to build, refine and apply three-dimensional general circulation models (GCMs). Such models include, but are not limited to, the NASA AMES model \citep{Pollack1993,Haberle2019} and the Laboratoire de Météorologie Dynamique planetary climate model (PCM; see \citealp{Forget1999,Millour2018}). Through the use of numerical models we are able to characterise Mars with limited observational data, allowing us to simulate areas where observational data are limited \citep{Read2015,Martinez2017}. Through these efforts our understanding of many atmospheric processes has been refined, including the annual CO$_2$ cycle \citep{Forget1998, Malin2001, Aharonson2004, Hayne2012, Holmes2018, Banfield2020}, CO$_2$ availability in the interest of terraforming \citep{Jakosky2018}, its hydrological cycle \citep{Houben1997,Haberle2001, Brown2014, Shaposhnikov2016, Shaposhnikov2018, Singh2018, Pal2019}, its surface topography \citep{Smith1999, Richardson2002a, Zalucha2010} and the effects of the climatically dominant dust cycle \citep{Kahre2010,Wang2015, Forget2017, Wang2018, Gebhardt2020, Ball2021, Chaffin2021}. Simulations have been performed ranging in scale from global \citep{Navarro2014, Streeter2020, Kass2020} to mesoscale levels \citep{Montabone2006, Spiga2009, Newman2021}.

There are, however, still processes which are difficult to capture in climate models. One of these is dynamically modelling annular shifts in CO$_2$ freezing and thawing and the subsequent change in surface pressure this causes \citep{Paige1992, Forget1999, Haberle2008, Kahre2010}. Simulating this in a self-consistent way using a GCM is difficult, and efforts so far have relied on parameterisations (described by \citealp{Forget1998, Forget1999, Spiga2017, Singh2018, Gary-Bicas2020}). Recent model developments by \citet{Way2017} have included the CO$_2$ effects dynamically, providing a promising avenue for the development of existing Martian GCMs. This is beneficial because it captures the secondary effects of CO$_2$ precipitation between the atmosphere and surface as it descends. Another major challenge these GCMs face is accurately underpinning the cause of inter-annual dust storms and characterising dust-uplifting rates prior to and following the dust season \citep{Mulholland2013, Spiga2013, Forget2017}. Parameterising the methods for dust uplifting has been essential for simulating the climate and weather of Mars but has still required periodic manual adjustments in order to match observations \citep{Montabone2020}. This limits the efficacy and self-sufficiency of Martian climate models, leading to difficulty in simulating global dust storms that should occur without forcing across multiple years and difficulty in predicting more local dust storms. As \citet{Way2017} have highlighted by dynamically solving for pressure variation as a consequence of CO$_2$ precipitation, a more complete physical model capable  of capturing dust processes should also have beneficial feedback for the system.

To work towards rectifying these gaps in our modelling capabilities, we have adapted the Unified Model (hereafter UM) to the study of the Martian climate as a foundation step in building a comprehensive Martian climate model complementary to existing modelling efforts. The UM is used routinely for Earth weather and climate modelling. It has also been used for other planetary climates, including Earth-like exoplanets \citep{Boutle2017, Sergeev2020, Eager2020}, hot Jupiters \citep{Mayne2014, Lines2018, Drummond2018} and mini Neptunes \citep{Mayne2019}. By adapting the existing and well-tested Earth parameterisations to Martian conditions, we can model climate processes in comparative ways to existing Mars GCMs (e.g. the quantity of available dust and size of the particles in the atmosphere). Such steps are key if GCMs are to progress to characterising Mars' climate with less manual prescription of parameters \citep{Forget2017, Montabone2020}. In this aspect, the UM is self-consistent, dynamically solving for dust availability and using that prognostically to simulate Martian dust content throughout the varying seasons. In this first study, we focus on a dry climate, including orography and dust, and highlight the importance of capturing dust accurately and its influences on the Martian climate. We also show that our adapted UM simulations capture key large-scale features of the Martian atmospheric circulation, including a periodic dust cycle without the prescription of fixed dust parameters.

In this paper, we present a description of the UM and the adaptations made for the Martian climate (Sect.~\ref{sec:Model_description}). That is, we describe the key model adaptations to simulate a dry Martian climate, such as planetary variables (atmospheric composition and orbital parameters, Sect.~\ref{subsec:orbital}), radiative transfer (Sect.~\ref{subsec:radi_trans}), orography (Sect.~\ref{subsec:orog}), dust (Sect.~\ref{subsec:dust}) and atmospheric surface pressure (Sect.~\ref{subsec:pres}). In Sect.~\ref{sec:setup} we describe how we configure the UM output for analysis, including a configuration for two scenarios where one features radiatively active Martian dust (RA dust) and one has radiatively inactive dust (RI dust). In Sect.~\ref{sec:results} we present results from the two configurations of the UM for Mars and validate the UM with RA dust against the PCM. We highlight how differences in dust parameterisation create differences between outputs, both within the UM scenarios and between the RA UM and PCM, and then discuss the reasons for these and their implications. Finally, in Sect.~\ref{sec:conclusion}, we discuss how the model can be further developed to improve its accuracy, mainly via the inclusion of schemes capturing the effects of CO$_2$ ice and water vapour planned in future work.

\section{Model description}
\label{sec:Model_description}

\begin{figure}[t]
    \includegraphics[width=7.9cm]{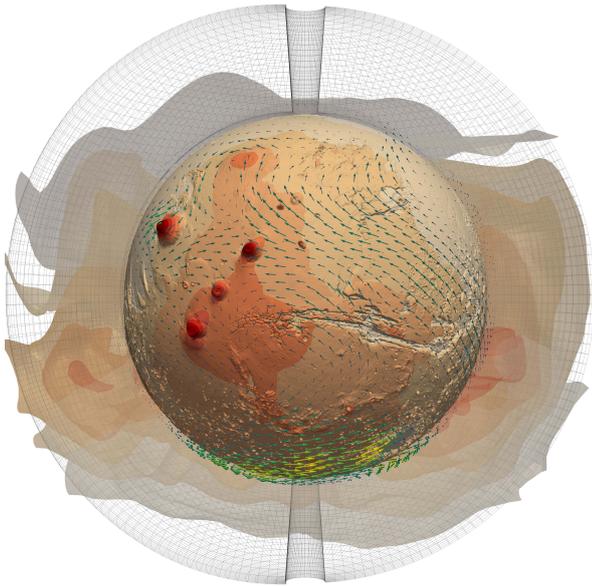}
    \caption{Overview three-dimensional plot of example output during Southern Hemisphere summer ($L_\mathrm{s}=260${\degree}). Included is a segment of the extracted regional dust layer as an isosurface, wind vectors at 1\,km height (arrows) and orography. Grid cells are cropped to 30\,km. Higher-resolution GIF and code are available at \url{https://github.com/dannymcculloch/3d_Mars_gif} (last access: 16~January~2023). Made with PyVista \citep{Sullivan2019}.}
    \label{fig:3dplots}
\end{figure}

For this study, we take the Global Atmosphere 7.0 science configuration of the UM \citep{Walters2019} and adapt it to Martian conditions. The UM dynamical core (ENDGame, described by \citealp{Wood2014}) simulates the atmosphere as a non-hydrostatic fully compressible fluid and its numerical formulation uses a semi-implicit timestep and semi-Lagrangian advection scheme \citep{Benacchio2016}. A full description of the model's dynamical core is given by \citet{Staniforth2003, Staniforth2008} and \citet{Wood2014}, with a global climate configuration further detailed by \citet{Walters2019}. The benefit of using existing schemes in a GCM instead of creating new ones is that they capture essential atmospheric physics and are verified under a variety of conditions (e.g. dust on exoplanets; \citealp{Boutle2020}). Figure~\ref{fig:3dplots} shows an illustrative example of output from the UM configuration used to simulate the Martian climate.

The UM grid configuration used in the present study has 90 by 144 grid points, corresponding to a resolution of 2{\degree} in latitude and 2.5{\degree} in longitude (i.e. a grid spacing of 118 and 147.5\,km at the equator, respectively). This resolution allows us to accurately capture climate trends at a relatively high spatial resolution \citep{Forget1999,Way2017}, with even higher spatial resolutions available at the cost of increased computational power. This resolution is suitable for observing seasonal patterns within the Martian year \citep{Navarro2014, Madeleine2011, Pottier2017a}, but higher spatial resolutions could readily be used to investigate selective regional climates (see for example \citealp{Sergeev2020}) -- a promising prospect for future applications.

In the vertical, we adopt 50 hybrid-height atmospheric levels up to a model top of 80\,km above the areoid level. We use a quadratically stretched grid to enhance resolution near the surface. Levels nearer the surface follow the terrain but are smoothed out gradually as the height increases, reaching a constant level height towards the highest altitudes (for our study this happens at $\sim $\,52\,km, \citealp{Wood2014}). The level heights can be seen in Table~\ref{tab:vertlevs}, where values are shown for a point with 0\,m surface height.

For this study we use planetary parameters (Sect.~\ref{subsec:orbital}), radiative transfer effects (Sect.~\ref{subsec:radi_trans}), orography (Sect.~\ref{subsec:orog}), prognostic dust (Sect.~\ref{subsec:dust}) and pressure (Sect.~\ref{subsec:pres}) set to Martian values, with more description of each process in the respective section.

\subsection{Planetary parameters}
\label{subsec:orbital}

Mars has an eccentric orbit ($e=0.0934$ compared to Earth's $e=0.0167$) leading to an annual oscillation in the received irradiation. To capture this in the UM, we configure the model to run with this orbit starting from $0{\degree} L_\mathrm{s}$ ($L_\mathrm{s}$ is the ecliptic longitude of the Sun),  and $0{\degree} L_\mathrm{s}$ corresponds to Northern Hemisphere (NH) spring equinox. We use the stellar output for the present-day Sun, but our simulated planet is placed at the Martian distance. Table~\ref{tab:marsorbit} shows the values we implement for Mars' planetary values.

\begin{table*}[t]
\caption{Orbital, planetary and atmospheric parameters in our simulations.}
\label{tab:marsorbit}
    \begin{tabular}{lr}
    \tophline
    Constant & Value \\
    \hhline
     Epoch (Julian date) & 2451545.0 \\
     Eccentricity & 0.0934 \\
     Obliquity (radian) & $\sim $\,0.4397 \\
     Mean acceleration due to gravity (m\,s$^{-2}$) & 3.711 \\
     Solar irradiance at 1\,AU (W\,m$^{-2}$) & 1361.0 \\
     Semi-major axis (AU) & 1.52368 \\
     Angular speed of planet rotation (radian\,s$^{-1}$) & $\sim 7.0882 \times 10^{-5}$ \\
     Radius (km) & 3389.5 \\
    \bottomhline
    \end{tabular}
\end{table*}

\subsection{Radiative transfer}
\label{subsec:radi_trans}

To calculate radiative transfer, we use the SOCRATES radiation scheme (described by \citealp{Edwards1996, Walters2019}). This scheme uses a two-stream correlated-$k$ method as described by \citet{Walters2019} and references therein. This scheme has been used extensively for studies of Earth \citep{Spafford2021}, in addition to hot Jupiters \citep{Mayne2014}, sub-Neptunes \citep{Drummond2018} and rocky exoplanets \citep{Boutle2020, Eager2020}.

For radiative properties, we use an adapted version of the ROCKE-3D spectral files\footnote{Files ``sp\_sw\_42\_dsa\_mars\_sun'' and ``sp\_lw\_17\_dsa\_mars'', available at  \url{https://portal.nccs.nasa.gov/GISS_modelE/ROCKE-3D/spectral_files/} (last access: 14~November~2022).}, which are appropriate for the CO$_2$-rich Martian atmosphere. We then added dust optical properties based on parameterisation from \citet{Walters2019}. Mars' atmosphere primarily consists of CO$_2$ ($\sim $\,95\,{\%}), N$_2$ (1.89\,{\%}) and Ar ($\sim $\,1.93\,{\%}) \citep{Read2015,Martinez2017}; we simplify this to 95\,{\%} CO$_2$ and 5\,{\%} N$_2$ in our simulations. Ar was omitted in this study as the effects would be minimal on seasonal averages. The prescribed gas ratios throughout the atmosphere are assumed to be well mixed \citep{Walters2019}.

The Martian atmosphere features small amounts of water vapour which are generally increased during the colder aphelion months \citep{Nazari-Sharabian2020}. This humidity affects the radiative transfer in every layer through water vapour molecules and cloud condensate  \citep{Shaposhnikov2016, Steele2017, Shaposhnikov2018, Fischer2019}. Mars has water ice clouds which influence radiative transfer between the surface and the upper atmosphere (e.g. \citealp{Navarro2014}). For our set-up, we use a completely dry atmosphere and surface. This is done to simplify the dust-uplifting processes and to be able to correctly capture Martian seasonal trends initially. This allows for a benchmark comparison which can be expanded on in future studies, with a similar approach being carried out by \citet{Turbet2021}.

In addition, we also use the terrain-shading scheme described by \citet{Manners2012}, which corrects the surface insolation depending on the zenith angle and obstructing elevation. This allows for a better representation of the effect that Martian orographical extremes have on their surroundings, e.g. the lone peak from Elysium Mons (25.02{\degree}\,N 147.21{\degree}\,E) casting a large shadow on the Northern Lowlands or the depths of the Valles Marineris canyon often being in a shade.

\subsection{Orography and surface}
\label{subsec:orog}

Orography affects various aspects of the Martian climate such as dust deposition and global circulation \citep{Smith1999, Zalucha2010, Pottier2017a}. Dominant orographic features include the Tharsis region and Hellas Basin but also a general hemispheric dichotomy featuring a higher southern hemisphere that gradually descends northward \citep{Richardson2002a}. Mars' hemispheric asymmetry heavily influences the atmospheric circulation, leading to large seasonal differences amplified by Mars' orbital eccentricity \citep{Richardson2002a, Zalucha2010}. Therefore, in order to better characterise Mars' climate and atmospheric processes, correctly capturing the surface elevation hemispheric dichotomy in Martian climate models is important \citep{Zalucha2010}.

\begin{figure*}[t]
    \includegraphics[width=15cm]{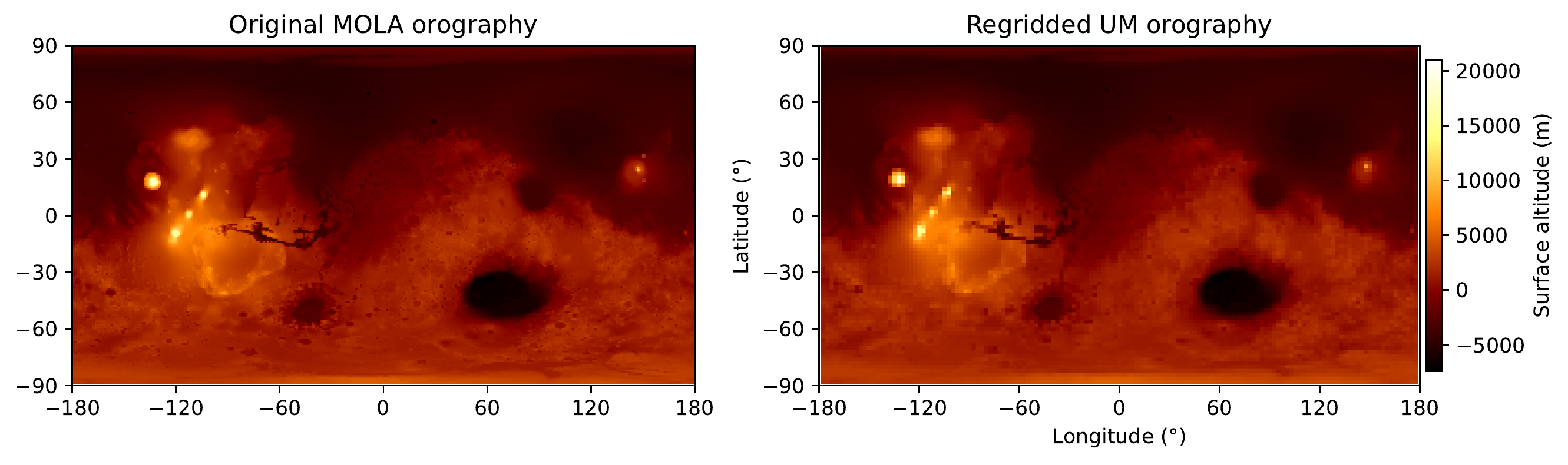}
    \caption{Original MOLA elevation data \textbf{(a)} compared to the regridded elevation data used in the UM \textbf{(b)}. Colour scales are matching between plots.}
    \label{fig:orog}
\end{figure*}

For this study, we use the sub-grid orographic drag parameterisation already present and verified in the UM (as described in detail by \citealp{Lott1997, Webster2003, Vosper2015a, Walters2019}) but for Martian values. This parameterisation allows for inter-grid-cell shading caused by areas of higher elevation (e.g. the upper edges of the Valles Marineris shading the crevice below). We obtained the widely used, high-resolution MOLA elevation data (Fig.~\ref{fig:orog}a, described by \citealp{Smith1999})\footnote{MOLA dataset available at \url{https://astrogeology.usgs.gov/search/map/Mars/Topography/HRSC_MOLA_Blend/Mars_HRSC_MOLA_BlendDEM_Global_200mp_v2} (last access: 15~June~2022).}. We regrid the MOLA dataset to the resolution used in the current study. We choose to use the resolution of $90\times 144$ as it allows for an adequate global representation needed to simulate global climate patterns present on Mars. In regridding, the cells from the original dataset that encompass a single grid cell are averaged, which leads to some height loss at the highest peaks, where sub-grid elevation is varied. The effects of the regridding can be seen in Fig.~\ref{fig:orog}b, where the MOLA dataset is compared to the regridded version. There is some inevitable height smoothing with regridding: Olympus Mons changes from 25\,km height to 19\,km and the lowest parts of the Hellas Basin from $-7.5$ to $-7.3$\,km.

For surface albedo, we assume a uniform value of 0.3. Although this is higher than the average albedo for Mars ($\sim $\,0.16 \citealp{Kieffer1977}), this was chosen as it compensates for the lack of polar CO$_2$ ice (which has an albedo of $\sim $\,0.5).
For surface thermal inertia, we use a constant value of 368.646\,J\,m$^{-2}$\,K$^{-1}$\,s$^{-0.5}$ across the surface (prescribed as a thermal capacity of $1.359\times 10^{-6}$\,J\,m$^3$\,K$^{-1}$), which is representative of the majority of the Martian surface \citep{Kieffer1977, Palluconi1981, Mellon2008}.

\subsection{Dust and surface roughness}
\label{subsec:dust}

Dust is synonymous with Mars: it is a key driver in a wide range of atmospheric phenomena, ranging from mesoscale dust devils, effective at uplifting local surface dust \citep{Neakrase2016}, to global dust storms affecting global temperatures for long periods of time \citep{Forget2017, Wang2018, Streeter2020}. Dust is a crucial contributor to the greenhouse effect on Mars. Because of this, large fluctuations in atmospheric dust content can have serious effects on the lower-altitude temperatures (below $\sim $\,25\,km) during global dust storms and have been well described by \citet{Streeter2020} and \citet{Wang2015}. Dust also affects the diurnal cycle of temperatures, retaining thermal radiation during the night and reflecting solar radiation during the day \citep{Madeleine2011}.
The dust quantities vary across the Martian year, with months 6 to 12 (month timings shown in Table~\ref{tab:months}) having much higher atmospheric dust than the other months \citep{Forget2017}. Months 1 to 6 are generally colder on Mars, leading to weaker wind speeds and subsequently less dust uplifting during the colder months. Intra-annual shifts from the dust storm season to a colder less dusty season are difficult to self-consistently capture in three-dimensional GCMs \citep{Madeleine2011, Forget2017}.

We adapt the dust scheme available in the UM, which handles dust parameterisation using nine particle radial size bins (0.03--1000\,\unit{\mu}m). A normalised distribution characterising the surface dust is set, with the values in bins 1--6 (0.03--30\,\unit{\mu}m) prescribed and the remaining dust equally distributed across bins 7--9 (30--1000\,\unit{\mu}m). This is described in detail by \citet{Marticorena1995} and \citet{Woodward2001, Woodward2011}, and an example of its application in a non-Earth climate can be seen in \citet{Boutle2020}. Atmospheric dust is absent upon initialisation and is calculated throughout the model simulation.
Dust particles are transported by atmospheric dynamics, turbulence \citep{Martin2000}, saltation (for uplifting larger particles, \citealp{Woodward2001, Woodward2022}) and dry deposition. Absorption and scattering of short-/long-wave radiation is calculated using Mie theory with the assumption that dust particles are spherical.

\begin{figure*}[t]
    \includegraphics[width=12cm]{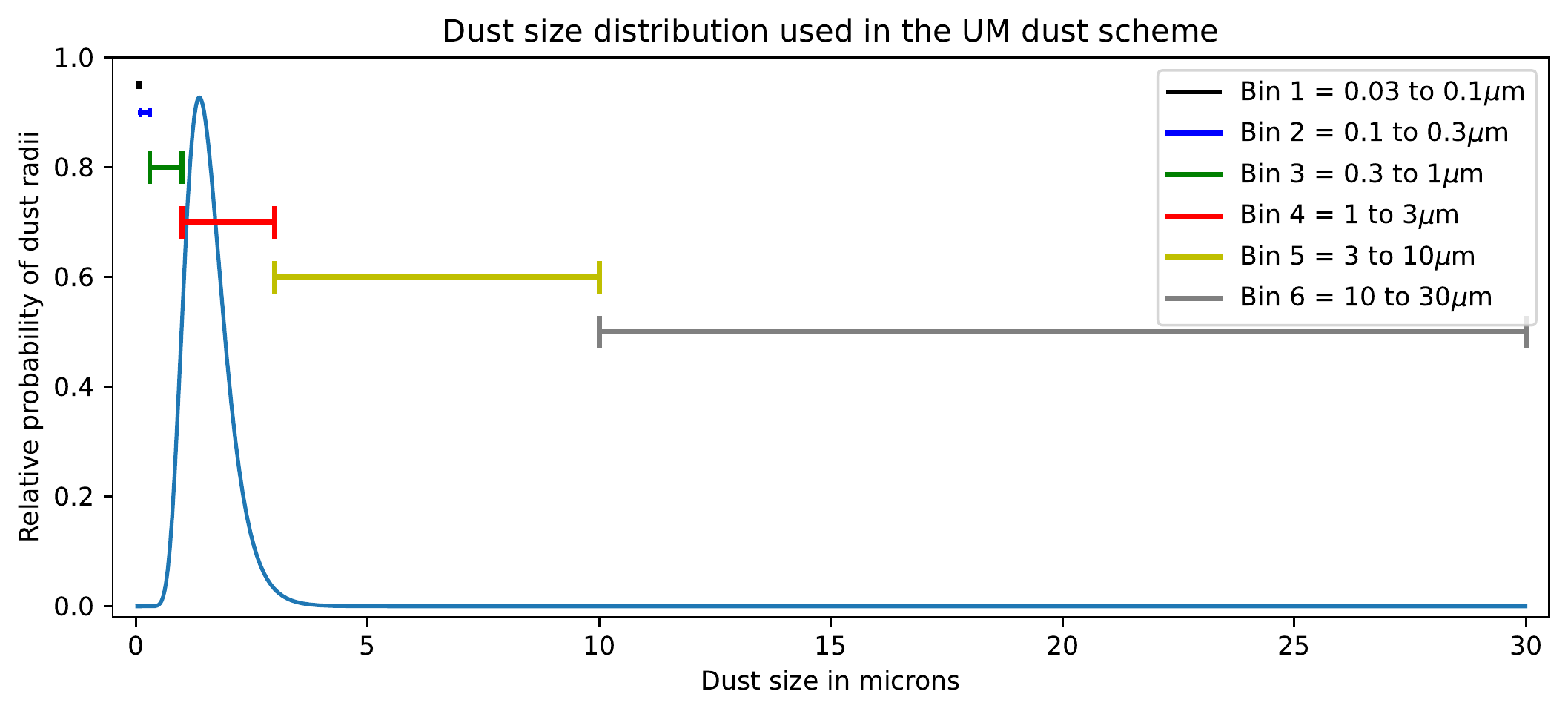}
    \caption{Dust size probability distribution used for the UM following \citet{Madeleine2011}. Dust bin ranges are shown by the coloured bars.}
    \label{fig:dustdist}
\end{figure*}

To determine the size distribution for the respective dust bins in the UM for Mars, we applied the same formula as used in the second scenario by \citet{Madeleine2011}, namely
    \begin{equation}
    \label{eq:dust_size}
        n(r) = \frac{N}{\sqrt{2\pi} \ \sigma_0 \ r} \exp{\left[-\frac{1}{2}\left(\frac{\ln(r/r_0)}{\sigma_0}\right)^2\right]},
    \end{equation}
where $r$ is the potential size of the dust particle, between 0.03 and 30 microns. $\sigma_0$ is the variance and $r_0$ is the mean, given by \citet{Madeleine2011} as 0.3 and 1.5\,\unit{\mu}m, respectively. $N$ is the normalised maximum (to unity) number of particles available and $n(r)$ is the probability of dust radii being present dependent on $r$. This provides the distribution presented in Fig.~\ref{fig:dustdist}, giving the probable radial size of any given particle. The dust bin size ranges used have been overlaid with coloured bars for each bin. The majority of dust resides in bin 4, but all bins are used in this study.

Dust production, uplifting and deposition are all characterised dynamically throughout the simulation. For dust production, upon initialisation, the surface is assumed to have an infinite amount of available dust to be uplifted. Dust production for a particle at rest is dictated by weight (primarily driven by particle size and composition), interparticle cohesion forces and wind shear stress along the surface \citep{Marticorena1995}. As the set-up is completely dry in the current configuration and we use a dry sand composition, the main factor which will impact dust production in this scenario is the particle size.

Dust uplifting is primarily driven by aeolian processes. Dust that has been freed from the surface is then transported via turbulent eddies (``suspension''), saltation and creeping. The ability for particles of a given size to be uplifted is proportional to their weight against the aerodynamic drag experienced \citep{Marticorena1995, Woodward2001}. Further dust uplifting as a consequence of saltation and creeping is influenced by the aerodynamic roughness length, which dictates how much dust is further uplifted via the impact of larger particles. This threshold value thus dictates how easy it is for smaller dust particles to be uplifted following an impact via saltation or creeping from a larger particle ($< 60$\,\unit{\mu}m in an Earth atmosphere; \citealp{Marticorena1995}). In the UM, horizontal dust flux is a tunable parameter. In this study, the horizontal dust flux was tuned to 7.5 to match atmospheric dust levels in the PCM for month 9. Initial testing of this parameter with different values did not change the distribution of the uplifted dust but solely changed the amount of uplifted dust in the localised regions.

\begin{figure*}[t]
    \includegraphics[width=15cm]{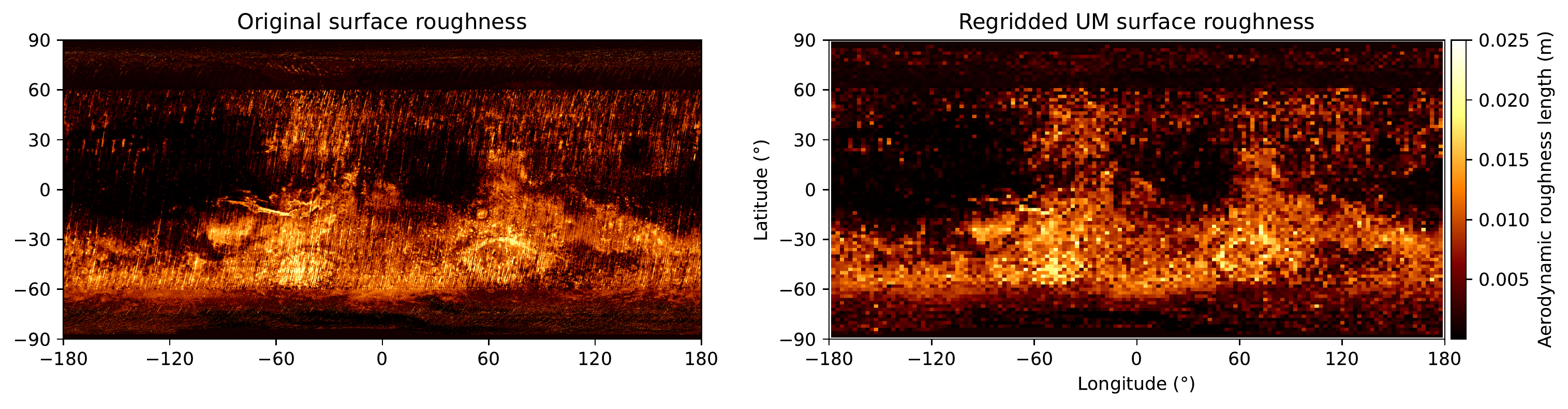}
    \caption{Surface roughness map from \citet{Hebrard2012} (left) and how it is represented in the UM after regridding (right).}
    \label{fig:surfaceroughness}
\end{figure*}

In the UM, we use the aerodynamic roughness length map from \citet{Hebrard2012}, shown in Fig.~\ref{fig:surfaceroughness}. This aerodynamic surface roughness threshold value could be set to a constant value in order to make sure that dust is being uplifted equally across the planet, but this is likely to oversimplify the climate. By using these values, we are able to simulate regional dust production dispersion, as opposed to a globally uniform dust production rate \citep{Hebrard2012}.

For dust optical properties, we used Earth dust optical values as shown in Table~\ref{tab:dustrefractindex} and described in \citet{Balkanski2007}. These values are primarily used for Earth dust, but as these values are highly similar to those used by \citet{Madeleine2011} and references therein, they were not changed.

\section{Experimental set-up}
\label{sec:setup}

We initialise the UM from a motionless atmospheric state with a uniform surface temperature of 250\,K and a surface pressure of 610\,Pa. Other variable schemes are also initialised at this stage (orography, Sect.~\ref{subsec:orog}, and dust, Sect.~\ref{subsec:dust}). The model is then integrated for 40 Martian years to achieve a steady state, which is determined as there being no inter-annual increase or decrease in the balance between incoming and outgoing radiation at the top of the atmosphere (TOA) between Martian years (with years averaged to omit for differences caused by orbit eccentricity). This also allows for localised prognostic dust reservoirs to form, with the model developing from a uniform surface dust amount (upon initialisation) to an atmosphere and surface with dust content as a consequence of the previous year's dust cycle. Thus, by the end of the 40-year period, the distribution of surface dust is non-uniform and features localised areas of high and low dust abundance. After this, the model is run for another Martian year, and this final year's data are what is presented in this paper. The model output is provided with every sol across a Martian year (668 sols/687\,d). A sol is defined here as 24\,h and 40\,min (simplified from a Martian solar day of 24\,h, 39\,min and 35\,s). For each model diagnostic, values are recorded every 5\,min (20\,min for radiation variables, e.g. TOA radiation flux) and are then averaged at the end of each sol. The year starting at 0{\degree} $L_\mathrm{s}$ is then run for 688 Earth days, taking an average across the sol. These sol outputs are then aggregated into Martian months (Table~\ref{tab:months}); this is done to better understand seasonal trends across the year and to match the data to time distributions present in other models \citep{Forget1999}.

To discern the effects of dust in the UM, we perform two separate simulations, one with radiatively active (RA) dust and one with radiatively inactive (RI) dust. Both set-ups are identical in every other way (e.g. spin-up time, orography, orbital parameters). For RI dust, the dust sizes and quantities are still prescribed, but all radiative effects of dust are switched off. This allows us to observe the effects of a dust scheme in our Mars set-up vs. what would already occur without the presence of atmospheric dust. This is useful for a variety of reasons. It allows us to highlight spatial/temporal regions of interest where dust might originate from, particularly where there are differences between scenarios. It also allows us to begin to distinguish the exclusive influence of dust, as any differences between scenarios can be attributed to this one variable.

To ensure that the UM reproduces seasonal patterns with sufficient accuracy, we compare our results to those of an established Mars GCM. In this study, we use year average (an average of all years where a dust storm did not occur) results from the Mars Climate Database\footnote{Available at \url{http://www-mars.lmd.jussieu.fr/} (last access: 6~March~2022).}, which provides output from the PCM \citep{Forget1999, Millour2018}. This dataset has 49 points in latitude and 64 points in longitude, with 30 layers in the vertical extending up to 108\,km. The output is also separated into the same Martian months as prescribed in the UM (e.g. Martian month 1\,$=$\,sols 0 to 61, as per Table~\ref{tab:months}). It features dust \citep{Madeleine2011}, a hydrological cycle \citep{Navarro2014}, a CO$_2$ ice cycle \citep{Forget1998, Forget1999}, atmospheric ozone \citep{Lefevre2008} and an upper atmosphere layer above 80\,km \citep{Colaitis2013a, Gonzalez-Galindo2015}.

The PCM uses a terrain-following pressure-based vertical coordinate $\sigma = p/p_\mathrm{s}$, where $p_\mathrm{s}$ and $p$ are the surface pressure and atmospheric pressure respectively, which is different to the UM's height-based vertical coordinate \citep{Forget1999, Wood2014}. This presents a difficulty in precise comparison between model simulations, as results cannot be compared straightforwardly without some form of interpolation. Furthermore, the UM and PCM have different upper boundaries, which require datasets to be cropped until the height is matched. Therefore, to validate our model against the PCM, we linearly interpolate the UM output to $\sigma$ levels at each output timestep, focusing on the levels where there is sufficient data for both models at the same pressure ($\sigma$ ranging from 1 at the surface to 0.01 in the upper atmosphere). As we are concerned with the large-scale seasonal climate we compare zonal, monthly averages of the UM outputs to those of the PCM.

\section{Results}
\label{sec:results}

In the following sections, we show Mars' annual mean pressure observations at the \textit{Viking} lander sites compared to the UM (Sect.~\ref{subsec:pres}). We then describe Mars' climate seasonality and how this is portrayed in simulations (Sect.~\ref{subsec:year_overview}). We then highlight the key differences between the RA dust scenario and PCM in more detail (Sect.~\ref{sec:PCM}). In particular, we focus on dust differences in our results and discuss further the implications of this (Sect.~\ref{subsec:dustdisc}).

\subsection{Atmospheric pressure}
\label{subsec:pres}

\begin{figure}[t]
    \includegraphics[width=8.3cm]{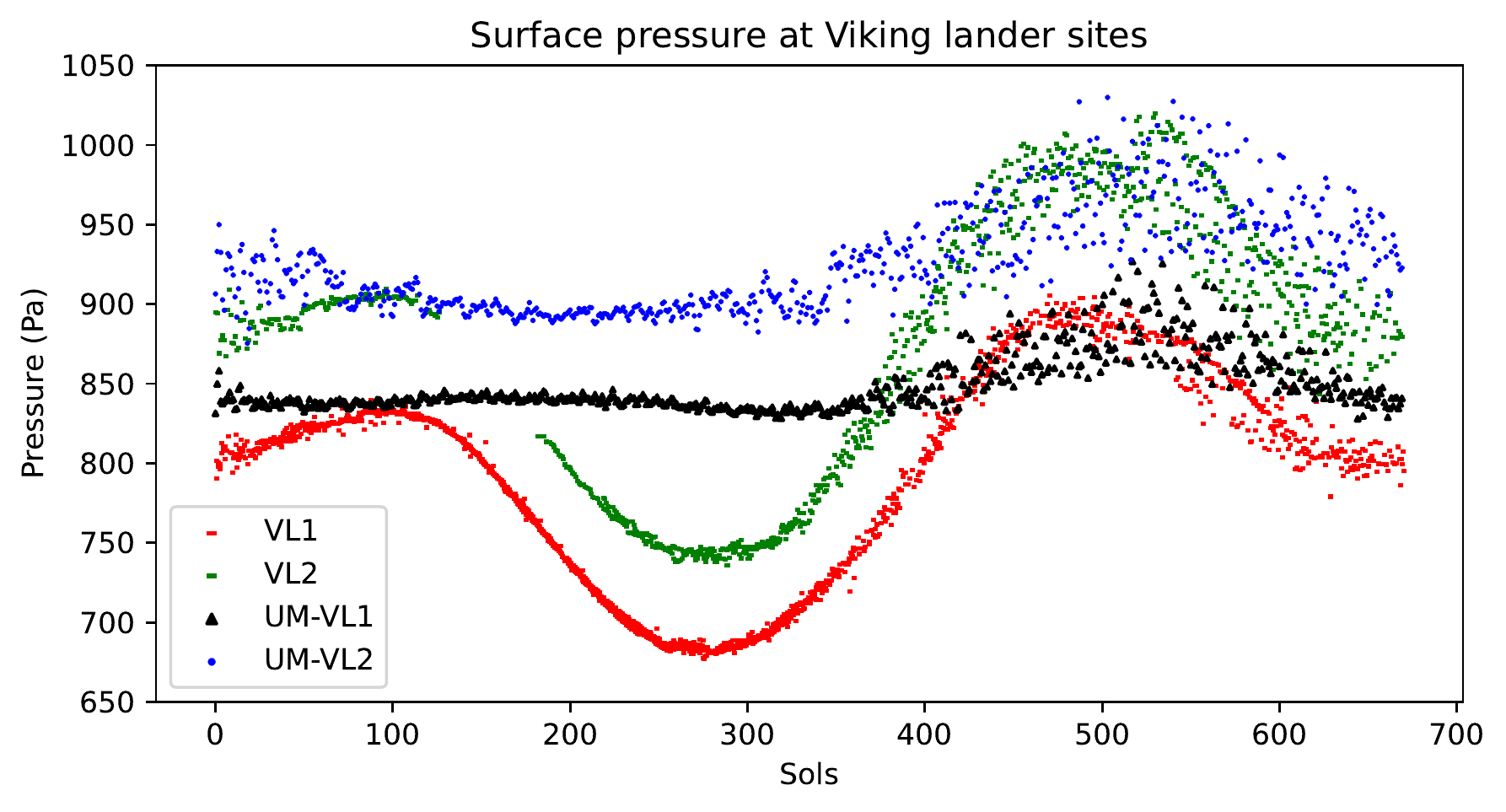}
    \caption{UM surface pressure at approximate \textit{Viking} lander 1 and \textit{Viking} lander 2 sites compared to observational data across a Martian year. \textit{Viking} lander data available from \citet{Tillman1989}.}
    \label{fig:pres_VL}
\end{figure}

Mars undergoes annual fluctuations in surface pressure, decreasing during colder months and increasing during the dust season. This is mainly due to the net freezing and thawing of polar ice caps, which extract and release atmospheric CO$_{2}$, respectively. The mean surface pressure on Mars is much lower than on Earth, resulting in large diurnal temperature fluctuations and limiting dust loading capacity \citep{Read2015, Martinez2017} due to less heat retention from the atmosphere. To show how pressure fluctuates in the UM, we show surface pressure across the Martian year at the approximate \textit{Viking} lander sites in the model compared to observational data from the landers. The values from the UM are not from the exact spatial location of the landers, due to their positions being within grid cells, but instead are the closest data points to where the landers would be in the UM. Figure~\ref{fig:pres_VL} shows that the UM-observed pressure near the \textit{Viking} lander 1 and 2 (VL1 and VL2) sites remains steady during the colder months and increases during the dust season. As the UM does not currently have a CO$_{2}$ ice cycle, this cold season pressure drop present in observations is absent. This leads to a high disparity of pressures between sols 130 and 410 between simulated and observed pressures. The UM does capture a minor pressure increase during sols 410--530 despite the lack of a CO$_2$ ice scheme. This increase in pressure is likely due to a temperature increase as a result of higher solar radiation. The absence of a pressure decrease during months 2--6 would suggest that this could potentially be a secondary feedback effect caused by heating from atmospheric dust. Since atmospheric dust quantities are much lower in the colder months, their effect is minimal, but as dust abundance increases, the magnitude of the effect of it is amplified, increasing pressure further.

\begin{figure*}[t]
    \includegraphics[width=17.5cm]{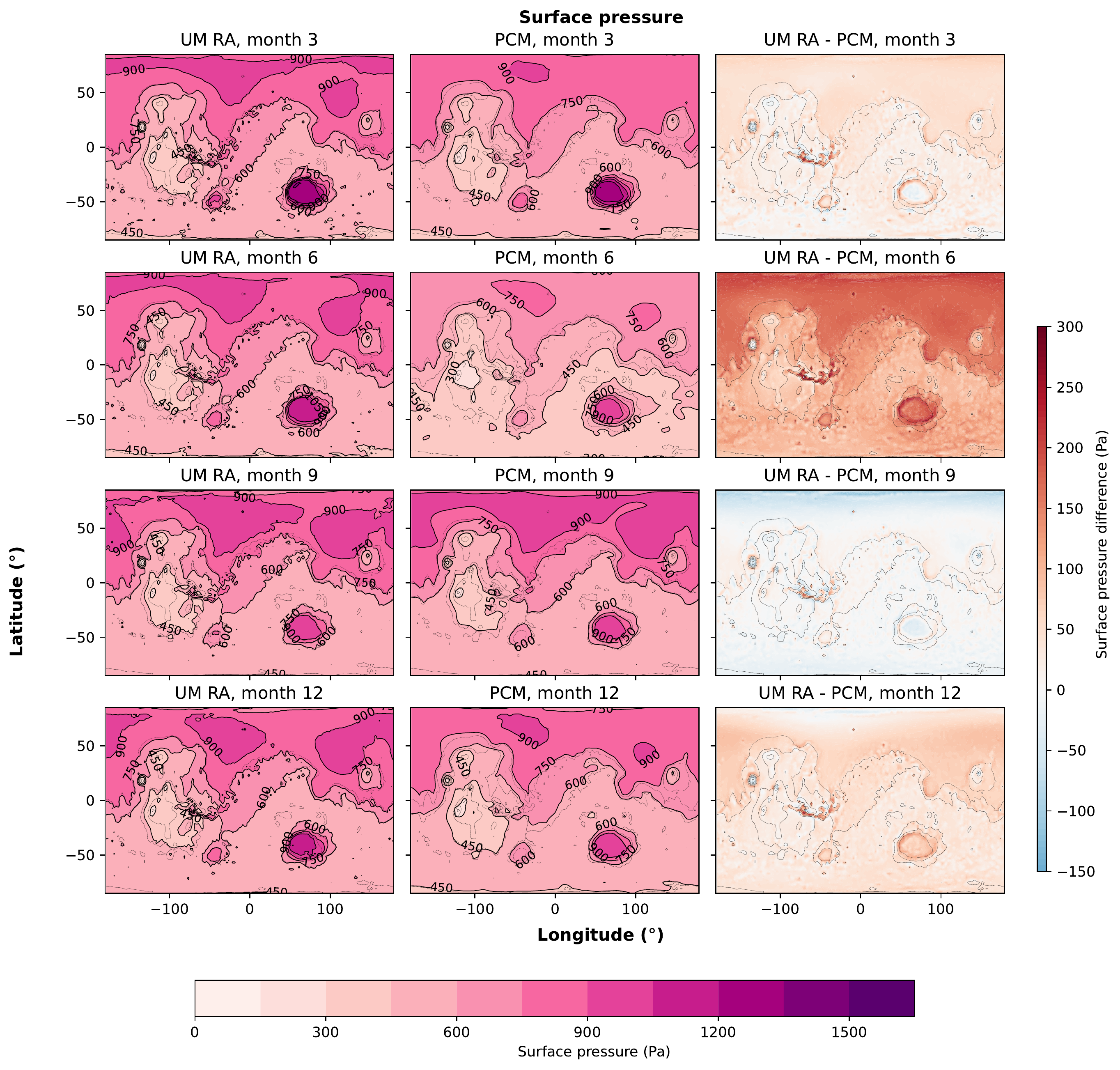}
    \caption{Surface pressure (Pa) across four seasons within the Martian year. For each month, the time average is taken of all sols within that month. The RA dust scenario is shown on the left, the PCM output is in the centre and the differences between the RA and PCM outputs are on the right. Colour scales in the left-hand and centre plots are matched across all months and between models with contour intervals of 150\,Pa. The colour scales in the difference plots are also matched across all months. Contours are not shown due to the sharp changes in pressure around craters creating steep differences within a small area.}
    \label{fig:surf_pres}
\end{figure*}

Accurate surface pressure is important for characterising the climate, as it affects processes such as the thermal capacity of the atmosphere and transport of material (such as dust) across hemispheres \citep{Gierasch1973, Hourdin1993, Hourdin1995, Read2015, Martinez2017}. Figure~\ref{fig:surf_pres} shows surface pressure across the year for the UM RA and PCM outputs. It shows that interaction observed in Fig.~\ref{fig:pres_VL} but as it occurs across the planet. In general, the UM RA features higher pressure than the PCM, especially in month 6 just after when atmospheric pressures are at their lowest (median sol of month 6 is sol 345). Surface pressures are most similar during month 9, with the majority of the planet featuring equal or slightly lower pressures than the PCM. There are, however, regions of extreme pressure difference in month 9. The UM possesses higher surface pressure in the depths of the Valles Marineris and lower pressure at the peak of Olympus Mons and at the NH pole. The causes of these localised pressure differences are likely the differences in orography used in the UM and PCM, with the UM having higher peaks and deeper channels.

\subsection{Overview of a year of Martian climate}
\label{subsec:year_overview}

To compare the results between RA dust, RI dust and the PCM, we show outputs of four atmospheric variables for months 3, 6, 9 and 12 (sol breakdown given in Table~\ref{tab:months}). Output is meaned zonally and temporally across the sols of the given month. The variables are zonal (eastward) winds (Fig.~\ref{fig:zonal_winds}), meridional (northward) winds (Fig.~\ref{fig:meri_winds}), temperature (Fig.~\ref{fig:temp}) and dust mass mixing ratio (MMR, Fig.~\ref{fig:dust}). The $3\times 4$ format and month dates are consistent throughout the figures\footnote{Additional figures are provided in the Appendix that compares only two models at once (i.e. RA--RI and RA--PCM); these plots allow for better one-to-one comparison between the RA output and RI/PCM outputs.}. We explain the development of Mars' climate across a typical year and how this is simulated in our results, and we then compare these outputs across simulations.

\subsubsection{Month 3: $L_\mathrm{s}$ 60--90{\degree}}

During this period, Mars is close to its coldest. NH temperature maxima are $\sim $\,220\,K at the northerly latitudes for both scenarios, zonal and meridional winds are slower and uplifted dust quantities are low. Temperatures drop due to Mars' orbit taking the planet away from the Sun, which leads to a reduction in solar radiation and a net cooling for the atmosphere and surface. This leads to a variety of secondary effects occurring on Mars. In reality, CO$_{2}$ begins to freeze more quickly than it thaws on the opposite pole, leading to a global pressure reduction as CO$_2$ is sequestered from the atmosphere, though this is not currently simulated. Temperatures vary throughout the Martian year, but averages are lowest during aphelion months (Fig.~\ref{fig:temp}, month 3) and highest during the dust season (Fig.~\ref{fig:temp}, month 9). Temperature oscillation caused by Mars' eccentric orbit does not seemingly have a direct impact on atmospheric pressure in the UM, as surface pressure remains consistent across aphelion months (sols $\sim $\,150 to $\sim $\,300). Instead, the only influence on atmospheric pressure in the UM (as seen in Fig.~\ref{fig:pres_VL}) is caused by the increase in dust abundance in the atmosphere during sols $\sim $\,360 to $\sim $\,660.  This temperature difference leads to weaker winds during month 3 and less dust uplifting, leading to less dust MMR throughout the majority of aphelion compared to months during perihelion \citep{Read2015}. Temperatures are highest in the NH, where it is summer, and then gradually decrease southward. Temperatures also decrease as height increases, as is to be expected. In terms of atmospheric circulation, Mars features strong zonal jets that alternate between hemispheres throughout the year, occurring during winter seasons to the respective hemisphere as the planet transitions between seasons (Fig.~\ref{fig:zonal_winds}). Meridional winds feature a weak jet near the equator surface, with an opposing jet at the upper boundary layer at $\sim 0.4 \sigma$ (Fig.~\ref{fig:meri_winds}). The UM RA dust and RI dust simulations are quite similar in month 3, with zonal wind differences of about $\sim $\,1--2\,m\,s$^{-1}$ and temperature differences of $\sim $\,1\,K. This is likely due to the low levels of dust abundance in both simulations, and thus its radiative impact during this time period is minimised (Fig.~\ref{fig:dust}, month 3).

The UM and the PCM both feature strong zonal jets in the upper atmosphere in the SH, but the UM's winds are slower at the upper boundary at the equator and in the NH zonal jet (Fig.~\ref{fig:zonal_winds}). The meridional jet in the UM is lower and slower than in the PCM simulations (by $\sim $\,4\,m\,s$^{-1}$, Fig.~\ref{fig:meri_winds}). Air temperature in the UM is generally lower, with especially strong differences between model outputs at the poles, negative at the south pole and positive at the north pole. The lower atmospheric temperature is likely caused by less dust in the atmosphere. Its radiative effects are minimal in the UM, whilst it is relatively abundant in the PCM (average temperature difference omitting the poles of $\approx -15$\,K, Fig.~\ref{fig:temp}). The temperature differences at the poles (exceeding $-30$ and 30\,K at either pole, Fig.~\ref{fig:surftemp}), which are much greater than those closer to the equator, are mainly due to the absence of polar ice in the UM, with the latent heat transfer and surface optical properties of ice being simulated in the PCM (Fig.~\ref{fig:temp}). Lastly, dust differences between model outputs are at their highest relative to concentration comparisons between models in this month. The UM simulates far less atmospheric dust than the PCM. This is likely due to the absence of forced dust uplifting and a dust devil parameterisation in the UM, which is present in the PCM (described by \citealp{Madeleine2011, Spiga2013, Montabone2020}).

\subsubsection{Month 6: $L_\mathrm{s}$ 150--180{\degree}}

\begin{figure*}[t]
    \includegraphics[width=14cm]{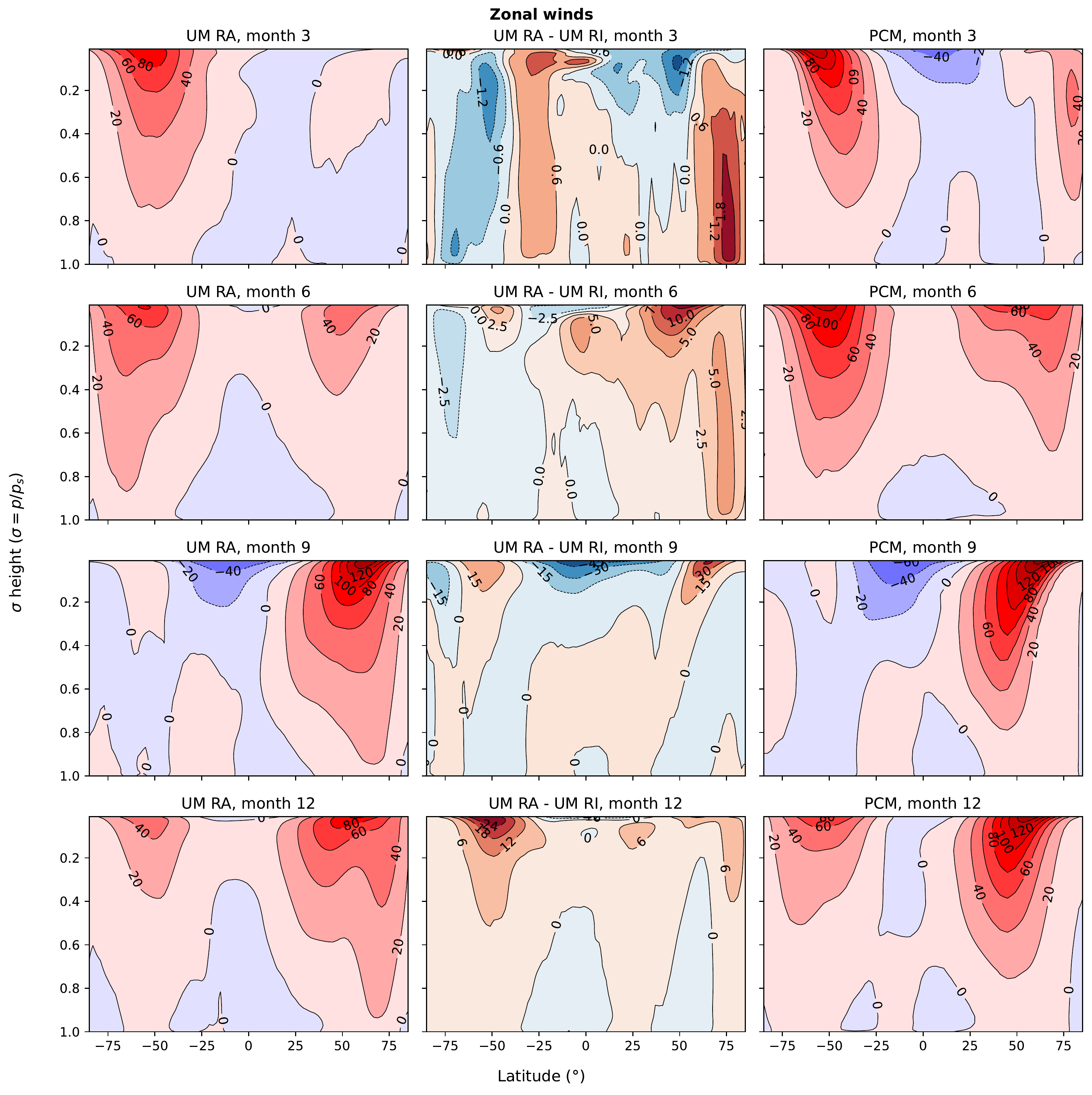}
    \caption{Zonal mean zonal winds (m\,s$^{-1}$) across four seasons within the Martian year. For each month, the time average is taken of all sols within that month. The RA dust scenario is shown on the left, the differences between the RA and RI dust scenarios are in the centre and the PCM output is on the right. Colour scales in the left-hand and right-hand plots are matched across all months and between models, with contour intervals of 20\,m\,s$^{-1}$. The contours in the difference plots are not matched due to the varying intensity of the difference between months. Positive values indicate an eastward wind and negative values a westward wind. Appendix Figs.~\ref{ap:RI_xwinds} and \ref{ap:PCM_xwinds} show the same data but solely for the RA vs. RI and PCM outputs, respectively.}
    \label{fig:zonal_winds}
\end{figure*}

\begin{figure*}[t]
    \includegraphics[width=14cm]{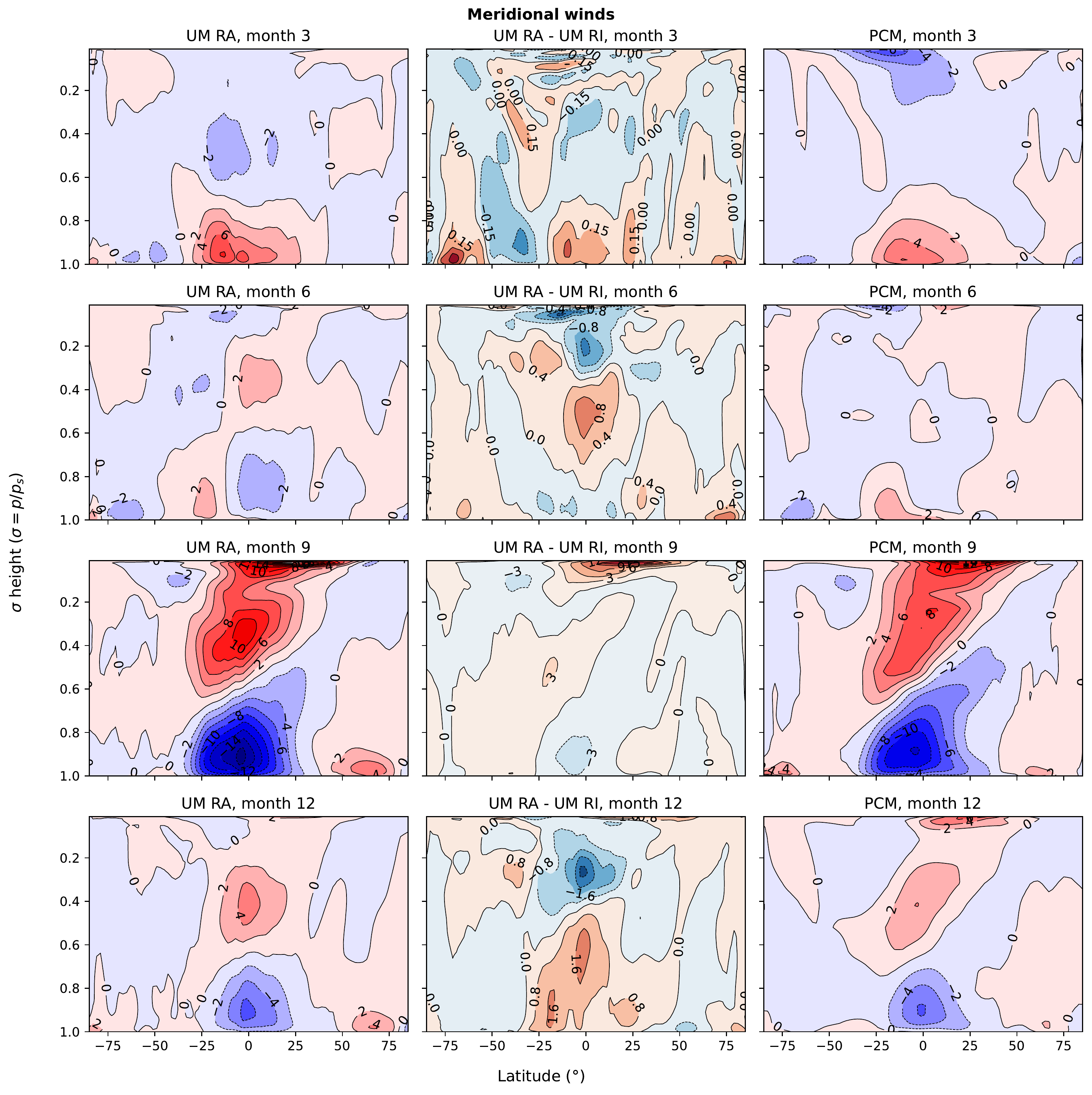}
    \caption{Zonal mean meridional winds (m\,s$^{-1}$) across four seasons within the Martian year. For each month, the time average is taken of all sols within that month. The RA dust scenario is shown on the left, the differences between the RA and RI dust scenarios are in the centre and the PCM output is on the right. Colour scales in the left-hand and right-hand plots are matched across all months and between models, with contour intervals of 2\,m\,s$^{-1}$. The contours in the difference plots are not matched due to the varying intensity of the difference between months. Positive values indicate a northward wind and negative values a southward wind. Appendix Figs.~\ref{ap:RI_ywinds} and \ref{ap:PCM_ywinds} show the same data but solely for the RA vs. RI and PCM outputs, respectively.}
    \label{fig:meri_winds}
\end{figure*}

\begin{figure*}[t]
    \includegraphics[width=14cm]{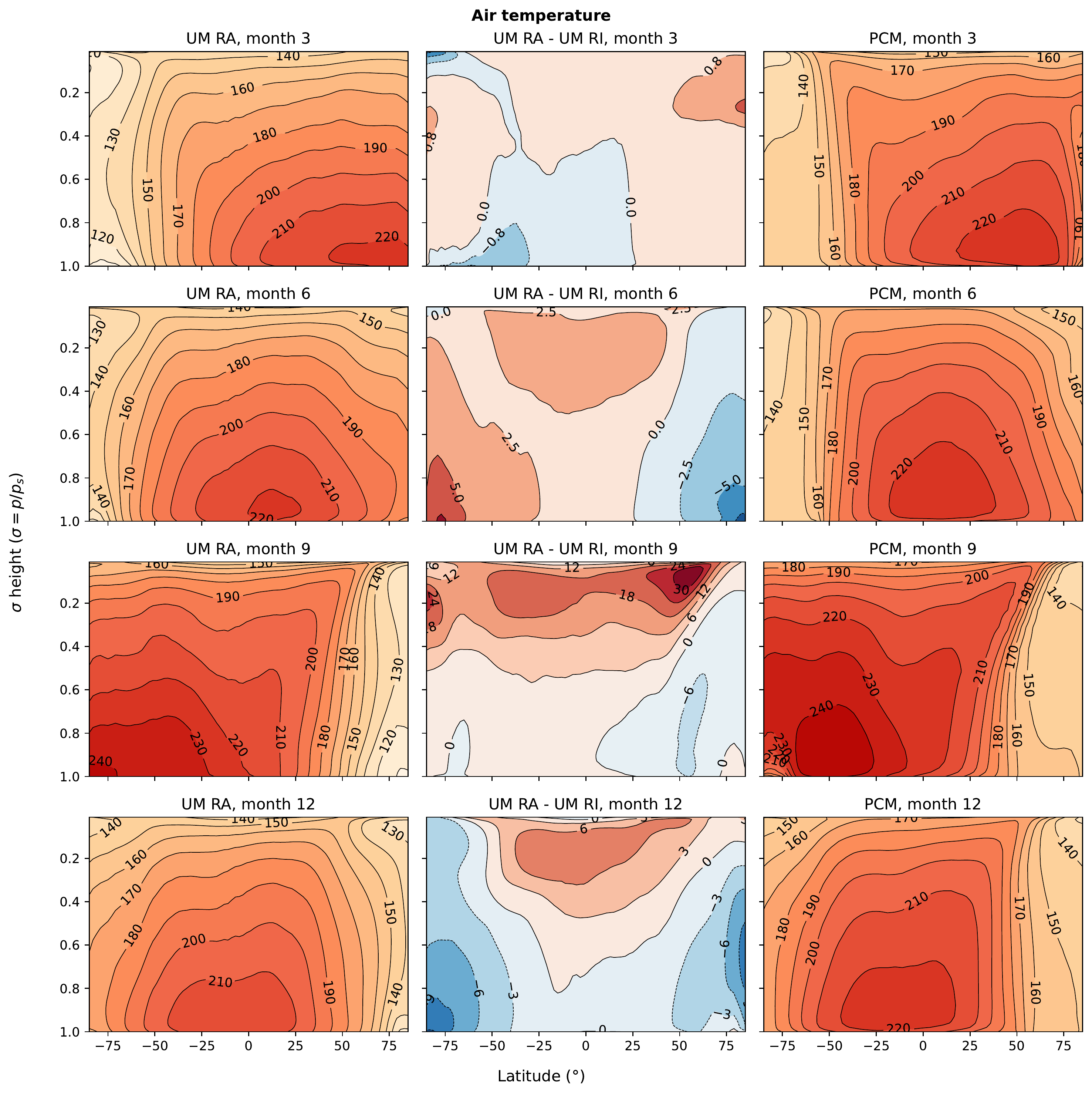}
    \caption{Zonal mean air temperature (K) across four seasons within the Martian year. For each month, the time average is taken of all sols within that month. The RA dust scenario is shown on the left, the differences between the RA and RI dust scenarios are in the centre and the PCM output is on the right. Colour scales in the left-hand and right-hand plots are matched across all months and between models, with contour intervals of 10\,K. The contours in the difference plots are not matched due to the varying intensity of the difference between months. Appendix Figs.~\ref{ap:RI_temp} and \ref{ap:PCM_temp} show the same data but solely for the RA vs. RI and PCM outputs, respectively.}
    \label{fig:temp}
\end{figure*}

\begin{figure*}[t]
    \includegraphics[width=14cm]{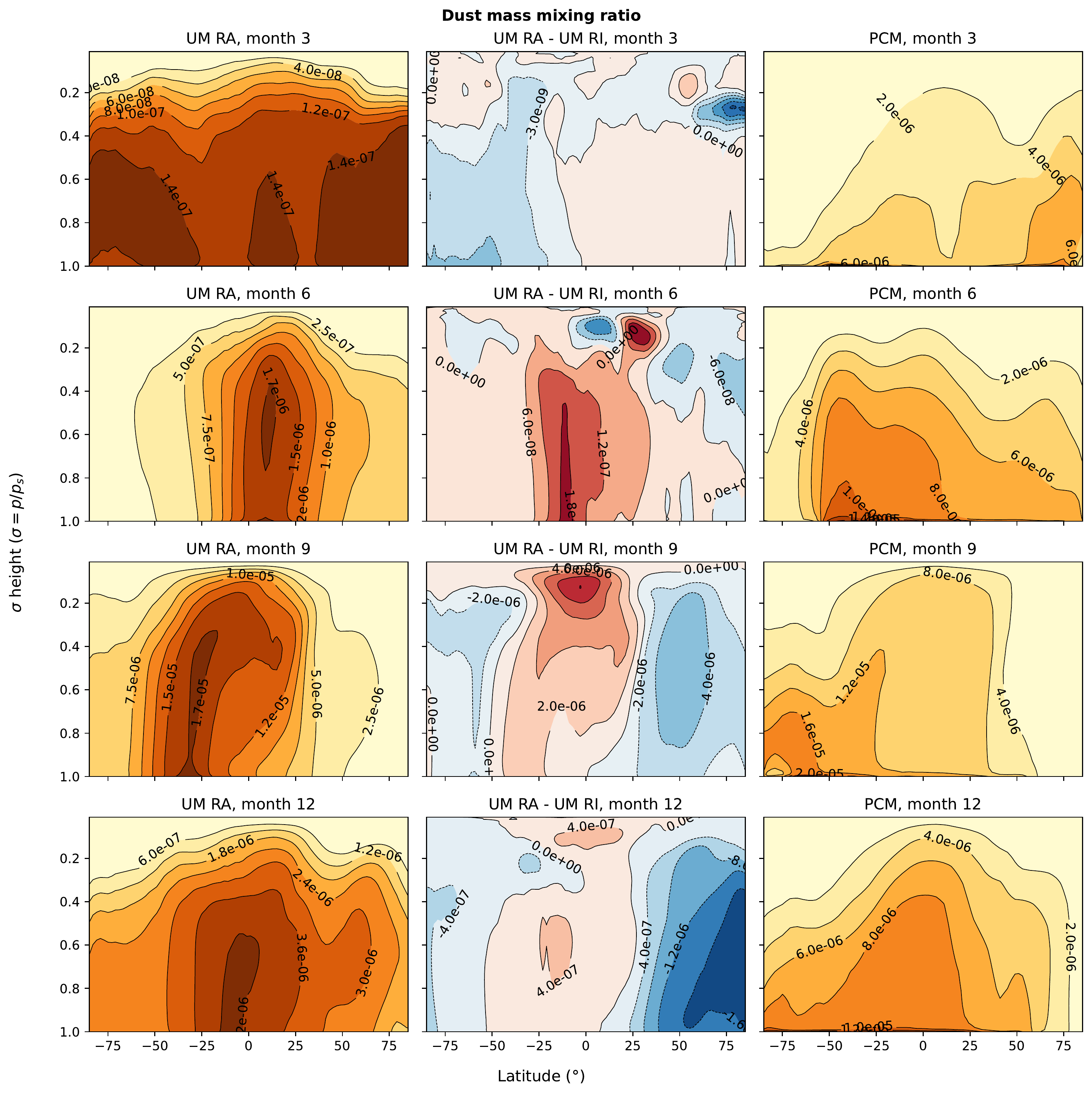}
    \caption{Zonal mean dust mass mixing ratio (kg\,kg$^{-1}$) across four seasons within the Martian year; each month is the average taken of all sols within that month (months according to Table~\ref{tab:months}). The RA dust scenario is shown on the left, the differences between the RA and RI dust scenarios are in the centre and the PCM output is on the right. Contour lines denote the mass mixing ratio and units are in kg\,kg$^{-1}$. Note that, due to the wide range of values present between months, the colour-scale ranges differ between months and models for this figure. Appendix Figs.~\ref{ap:RI_dust} and \ref{ap:PCM_dust} show the same data but solely for the RA vs. RI and PCM outputs, respectively.}
    \label{fig:dust}
\end{figure*}

Here, Mars' hemispheres are transitioning seasons; this can be seen in the jet reversal in the zonal winds and by the location of the temperature maxima at the lower latitudes. Mars features a single Hadley cell which reverses twice a year, with polar cells at each pole. Dust MMR is more than month 3 due to rising temperatures increasing wind speeds, with larger concentrations near the equator (Fig.~\ref{fig:dust}). Despite this, dust in the RA scenario is uplifted more compared to RI by $\sim $\,10\,{\%}. The reason for this difference is likely uplifted RA dust scattering solar radiation close to the surface, causing more near-surface warming than there would be with RI dust, where the increased temperature causes faster near-surface winds increasing dust-uplifting rates. Temperature differences between the RA and RI scenarios (Fig.~\ref{fig:temp}) are between $-5$ and 5\,K, which is more than in month 3 but lower than in later months. There is a clear difference between polar regions in opposing directions (colder NH pole and warmer SH pole), in addition to a mid-altitude band of warmer air by up to 3\,K in the RA dust scenario. These indicate that dust is beginning to impact the atmosphere more actively. Zonal winds are also reflecting the increasing differences (Fig.~\ref{fig:zonal_winds}), with an NH polar zonal jet difference of up to 40\,m\,s$^{-1}$. Meridional wind differences are still minimal (Fig.~\ref{fig:meri_winds}). Mars' seasonal cycle and the associated reversal of its Hadley cell during this month are clearly shown in the meridional wind patterns, with counter-flowing jets present in months 6 but more stable meridional winds during months 3, 9 and 12.

Differences between the UM RA dust scenario and PCM here are varied in their magnitude but are present for each variable. For air temperatures (Fig.~\ref{fig:temp}), the UM is comparable at the surface near the equator, but the temperature in the UM decreases more quickly with height up to the upper atmosphere. The UM, however, features a region with higher temperatures at $\sim $\,60{\degree}\,S latitude, which reaches $\sim $\,24\,K at the surface but gradually decreases with height. There are still differences between the models at the poles, but these differences are less substantial than they were during month 3 (now down to a difference of the UM being $-16$\,K compared to the PCM at the surface). The UM RA features a band of warmer air at $\sim $\,55{\degree}\,S by $\sim $\,20\,K (Fig.~\ref{fig:surftemp}), but temperatures near the SH pole are colder than in the PCM. Zonal winds feature a variety of differences between models, with faster and slower zonally averaged wind speeds distributed across the atmosphere. Both models feature polar jets in both hemispheres, but the UM zonal jets are slower than the PCMs by up to 40\,m\,s$^{-1}$ at the centre of the SH jet, which gradually becomes more comparable further from the centre of the jet towards the surface. There is an area of faster winds by $\sim $\,10\,m\,s$^{-1}$ in the UM at the SH pole, which is at the middle of the atmosphere and lessens towards the surface where differences again become 0. This is likely due to large polar temperature differences, leading to a weaker temperature gradient and causing the SH polar jet in the UM to be more stretched than the PCM. The polar jet in the NH is also slower in the UM by up to 30\,m\,s$^{-1}$. Dust differences between models are still considerable throughout the entire atmosphere as in month 3, with the PCM having up to $1.2\times 10^{-5}$\,kg\,kg$^{1}$ ($\sim $\,700\,{\%}) more dust than the UM near the surface, while differences lessen with height. Here, the UM dust quantities are much less than the PCM, but uplifting is increasing as Mars approaches the dust season. In all three scenarios dust concentration is increasing, this is promising as it shows the ability to dynamically simulate a substantial change in atmospheric dust abundance during a transition of seasons, as occurs on Mars.

\subsubsection{Month 9: $L_\mathrm{s}$ 240--270{\degree}}

\begin{figure*}[t]
    \includegraphics[width=14cm]{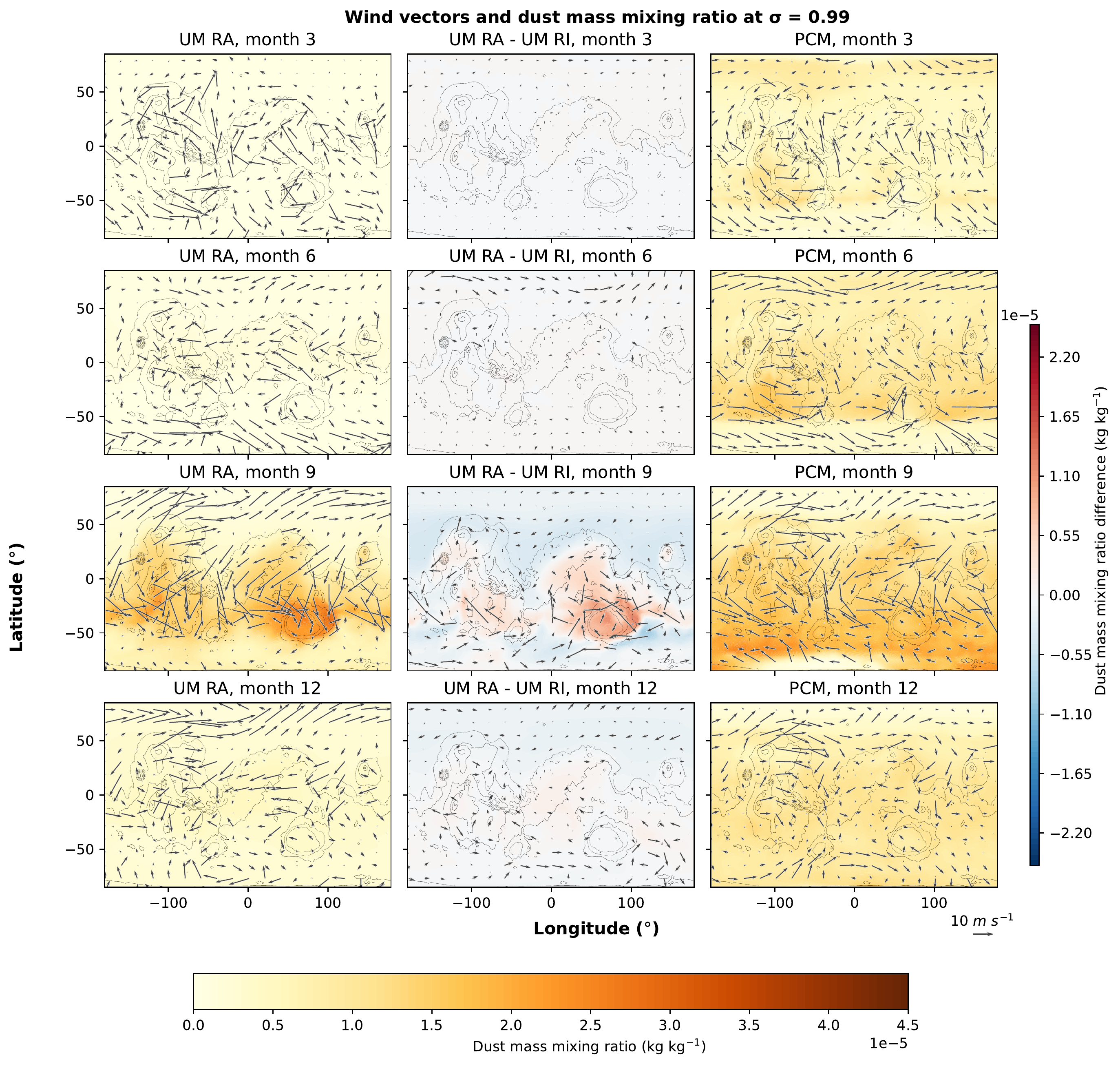}
    \caption{Dust mass mixing ratio (kg\,kg$^{-1}$) and horizontal wind speed (m\,s$^{-1}$) at $\sigma=0.99$ across four seasons within the Martian year; each month is the average taken of all sols within that month (months according to Table~\ref{tab:months}). The RA dust scenario is shown on the left, the differences between the RA and RI dust scenarios are in the centre and the PCM output is on the right. Colour scales in the left-hand and right-hand plots are matched across all months and between models, with contour intervals of $5\times 10^{-7}$\,kg\,kg$^{-1}$. The contours in the difference plots are not matched due to the varying intensity of the difference between months. The PCM output is shown in the centre of this figure to allow easier visual comparison between the UM RA dust scenario and PCM.   Appendix Figs.~\ref{ap:dustsurf_RI} and \ref{ap:dustsurf_PCM} show the same data but solely for the RA vs. RI and PCM outputs, respectively.}
    \label{fig:dustsurf}
\end{figure*}

\begin{figure*}[t]
    \includegraphics[width=14cm]{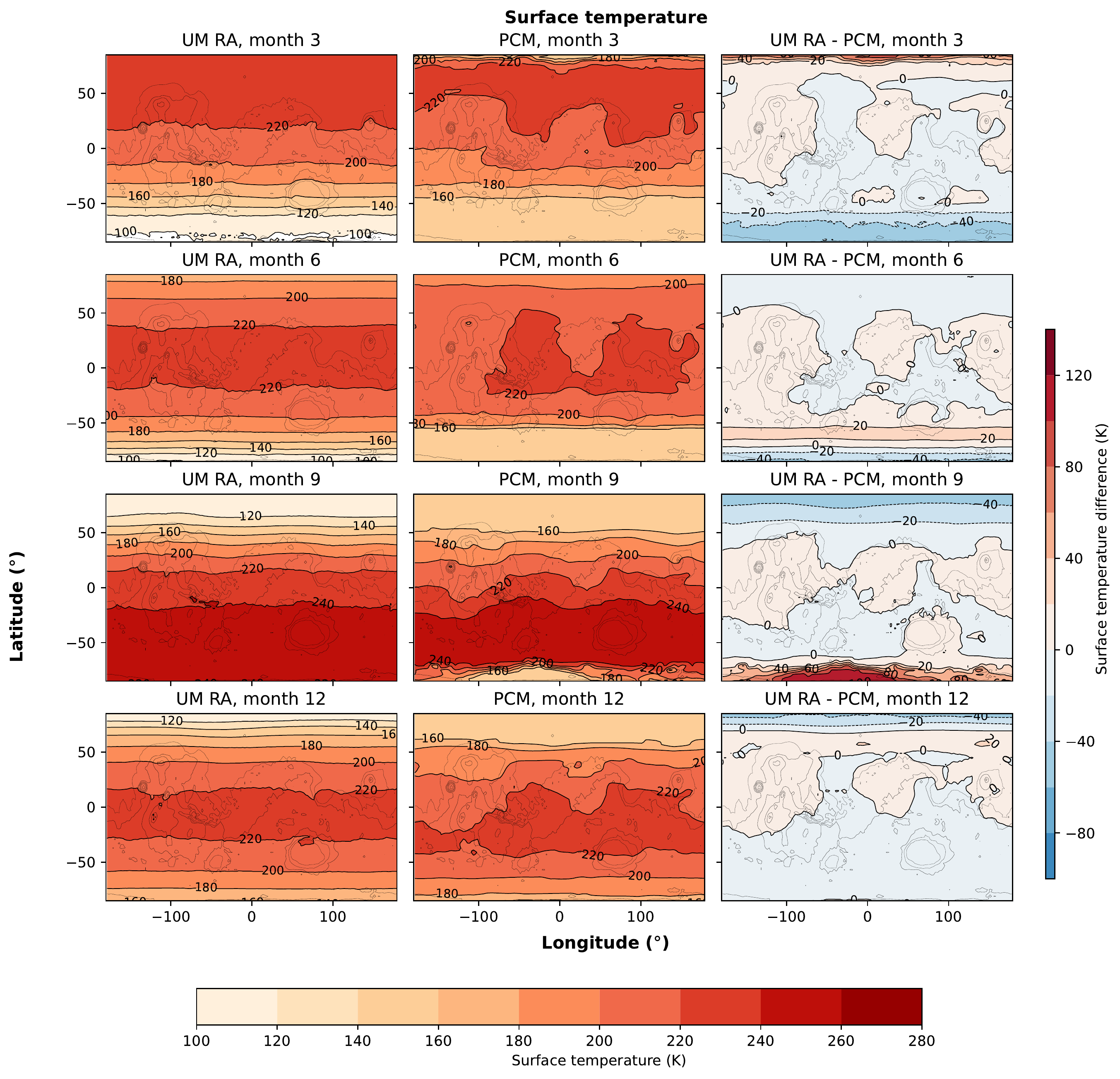}
    \caption{Surface temperature (K) across four seasons within the Martian year; each month is the average taken of all sols within that month (months according to Table~\ref{tab:months}). The RA dust scenario is shown on the left, the PCM output is in the centre and the differences between the RA and PCM outputs are on the right. Colour scales in the left-hand and centre plots are matched across all months and between models with contour intervals of 20\,K. The colour scales in the difference plots are also matched across all months and contour intervals are also 20\,K.}
    \label{fig:surftemp}
\end{figure*}

Month 9 is the peak of the dust season, and this is where differences between the RA and RI scenarios are greatest, as higher abundances of dust affect radiation fluxes more severely. Firstly, for both UM scenarios, dust (Fig.~\ref{fig:dust}, month 9) is mainly concentrated in the SH, forming a large ``plume'' that extends vertically. Dust abundances have increased from earlier months and are now 2 orders of magnitude more than in month 3. The dust in the RA simulation is more concentrated at $\sim $\,30{\degree}\,S latitude, while RI dust is distributed more widely across latitudes but with a minor difference towards $\sim $\,30{\degree}\,S latitude. This difference can clearly be seen with higher dust abundance at the equator at all altitudes for RA dust compared to less dust nearer both poles than RI dust. During colder months the temperatures are similar between UM scenarios, but during the dust season, the temperature differences are much more pronounced (Fig.~\ref{fig:temp}, month 9). This change in magnitude of differences shows the effects of RA dust on temperature, highlighting how dust influences radiative transfer in Mars' atmosphere -- particularly at mid altitudes, where dust can remain suspended, influencing incoming solar radiation and outgoing thermal radiation from the surface. Temperatures (Fig.~\ref{fig:temp}, month 9) near the surface are similar between scenarios, both being $\sim $\,20\,K warmer than month 3, but at lower pressures ($\sigma$ between $\sim $\,0.2 and 0.1), the RA dust scenario is considerably warmer (exceeding 30\,K). The RA dust scenario also features a NH pole which is $\sim $\,40\,K colder than the PCM but warmer by up to $\sim $\,80\,K at the SH pole. This highlights the radiative effects of suspended atmospheric dust and how the climate might be different in its absence. Suspended atmospheric dust causes the upper atmosphere to be warmer than if there were no effects from dust (Fig.~\ref{fig:dust}). In the scenario with RI dust, this dust layer does not affect incoming radiation, and solar radiation can reach the surface; however, in the RA dust scenario, the suspended dust layer scatters incoming solar radiation, transferring energy to the suspended dust layer instead of the surface. This ``band'' of warmer air stretches from the mid altitudes up to the top of the atmosphere, across almost all latitudes (with an exception at the NH pole). Faster wind speeds are a consequence of sharper temperature gradients due to the thermal wind balance relationship and so are affected by the disparity of temperature maxima between hemispheres. This can be seen by the higher wind speeds present in month 9 compared to month 3 (first columns in Figs.~\ref{fig:zonal_winds} and \ref{fig:meri_winds}). Zonal winds (Fig.~\ref{fig:zonal_winds}, month 9) feature more extreme differences between RA and RI simulations, ranging from $-40$ to 40\,m\,s$^{-1}$ in some places. These are mainly at the higher altitudes where temperature differences were at their greatest, above the dust layer. The polar jet in the RA dust scenario is considerably quicker but lacks an opposing jet near the top of the atmosphere in the SH ($\sigma = 0.1$, 50{\degree}\,S latitude), though this jet is small in the RI dust scenario ($-20$\,m\,s$^{-1}$). RI dust does not feature an equatorial jet in the uppermost part of the simulated atmosphere ($\sigma \leq 0.1$). Meridional wind differences for this month (Fig.~\ref{fig:meri_winds}, month 9) are at their highest compared to other months, ranging from $-3$ to 3\,m\,s$^{-1}$ in the lower/middle atmosphere ($\sigma = 1$ to 0.5, 15{\degree}\,N latitude). A region near the top of the atmosphere is faster in the RA dust scenario (up to 15\,m\,s$^{-1}$, $\sigma = 0.1$, 15{\degree}\,N latitude). The dust content in the RA scenario is more centralised around the equator, while dust content in the RI scenario is more spread across the planet, leading to more dust at the poles (Figs.~\ref{fig:dust} and \ref{fig:dustsurf}). This highlights the thermal feedback effects of atmospheric dust, as the dust content in the RA scenario causes more localised warming, driving increased vertical uplifting. This thermal influence is not present in the RI scenario, causing vertical uplifting to higher levels to be reduced, leading to a lower, more latitudinally dispersed, atmospheric dust layer. This shows the ability to simulate strong vertical wind-driven dust uplifting in the UM.

Differences between the UM RA dust scenario and PCM during month 9 are at their least compared to other months (when compared at the relative ranges of the different months), suggesting the importance of radiatively active dust for reproducing the salient features of atmospheric dynamics on Mars. Dust MMR in both models is now much more comparable, with the UM having uplifted more dust than the PCM during prior months. Spatially, dust in the UM is concentrated in a large central plume at $\sim $\,30{\degree}\,S latitude, with dust in the PCM being more spread out across the atmosphere. The reasons for this are uncertain but could potentially be the parameterisation of dust uplifting: the UM dynamically calculates dust reservoirs and horizontal flux, whereas PCM uses ``forced'' dust injection to more closely match observations \citep{Spiga2013, Montabone2015, Montabone2020}. This high vertical uplifting in the UM is responsible for a high input of dust into the atmosphere past the near surface, a process described in detail by \citet{Spiga2013}. The UM features comparable near-surface dust to the PCM across the $\sim $\,30{\degree}\,S latitudinal band (Fig.~\ref{fig:dustsurf}), but near-surface dust levels in the PCM are higher above and below this region (above 0{\degree}\,N and below  $\sim $\,50{\degree}\,S). Particular regions of higher near-surface dust in the UM are the Hellas Basin and the Tharsis region. Temperature differences between models exist throughout the atmosphere, with the largest differences occurring close to the surface at the poles and reaching $-40$\,K in the NH pole and 40\,K in the SH pole. These large differences are due to the current absence of ice at the poles in the UM, which will impact temperatures through emissivity and latent heat effects. The rest of the atmosphere is colder in the UM, with differences exceeding 20\,K above the SH pole and above the equator at the upper edge of the simulated atmosphere.  There is an agreement between models at the surface across the equator as in months 3 and 6, with another small area of agreement between models above the surface at $\sigma = 0.9$ to 0.2 and 50{\degree}\,N latitude. Zonal mean differences here are concentrated in the NH polar jet, with the rest of the atmosphere in agreement. The zonal NH polar jet in the UM is more spread out meridionally than the PCM, so the difference of 40\,m\,s$^{-1}$ at $\sigma = 0.2$, $\sim $\,60{\degree}\,N latitude is a result of the jet expanding horizontally in the UM rather than vertically as in the PCM. This is further highlighted by the $-20$\,m\,s$^{-1}$ difference at $\sigma = 0.3$, $\sim $\,45{\degree}\,N latitude, where the wind speeds in the UM are slower as a result of being more dispersed horizontally. Lastly, meridional wind differences between the UM and PCM in month 9 are not concentrated in a single jet or in a single direction but instead occur sporadically throughout the atmosphere. There are some differences of $-3$\,m\,s$^{-1}$ near the surface and below the top of the simulated atmosphere, with 3\,m\,s$^{-1}$ faster wind speeds around the middle atmosphere ($\sigma = 0.4$, 0 and 20{\degree}\,S latitude) and top of the simulated atmosphere ($\sigma \leq 0.1$, 0 to 40{\degree}\,N latitude).

When discussing dust, month 9 is the most relevant of the four selected months, as it features the height of the dust season when the dust is at its most abundant in both models. Our results show that we are able to simulate a dust cycle with intra-annual fluctuations in the UM without dust forcing. This month highlights the key takeaway from this study, that we have intra-annual dust quantity oscillation that is entirely reproduced dynamically by the GCM, with dust quantities rising during earlier months, peaking during the dust storm season and subsiding in later months.

\subsubsection{Month 12: $L_\mathrm{s}$ 330--360{\degree}}

In month 12, the dust season ends and the atmosphere of Mars cools. The winds are weaker and dust MMR subsides while the polar jets transition between hemispheres. The uplifted dust quantities are considerably less than those of the peak dust season, with RA scenarios still centred around the equator and RI dust shifting northward. Dust at the equator is more abundant in the RA scenario by $\sim 4.0\times 10^{-7}$\,kg\,kg$^{-1}$ ($\sim $\,10\,{\%} more dust in the RA scenario).

Overall, there is more dust uplifted in the RI scenario compared to the RA scenario, though this varies spatially (Fig.~\ref{fig:dust}, month 12). Dust content in the RA scenario at the NH pole is less than that in the RI scenario by up to $1.6\times 10^{-6}$\,kg\,kg$^{-1}$ ($\sim $\,187\,{\%}), with a similar but less extreme difference at the SH pole. Temperatures begin to decrease in this season, and the warmest region is again in the lower latitudes in both scenarios (Fig.~\ref{fig:temp}, month 12). The RI case is warmer at the equator in the upper atmosphere and colder near the poles in the lower atmosphere than that in the RA dust case. Zonal wind patterns change between hemispheres, with higher polar jet speeds in both hemispheres for RA dust (Fig.~\ref{fig:zonal_winds}, month 12). Differences in months 6 and 12 do switch hemispheres, but the magnitude of these differences is larger for month 12. This is likely caused by more residual dust in the atmosphere from month 9, causing temperature differences to be higher in month 12 than in month 6, in turn affecting the thermal wind relationship. This hemispheric reversal in the zonal wind mirrors what occurs in month 6 but is stronger in month 12 and stronger in the RA dust scenario, with maximum differences reaching now $\sim $\,20\,m\,s$^{-1}$ (compared to $\sim $\,10\,m\,s$^{-1}$ in month 6). Meridional winds for both scenarios show the Hadley cell direction reversal with the seasonal cycle (detailed by \citealp{Read2015}); wind speeds in both scenarios are once again comparable with the largest differences being between $-3$ and 1\,m\,s$^{-1}$.

Comparison between the UM RA dust scenario and PCM shows similar trends to month 6 in each variable but inverted with respect to latitude (i.e. warmer temperature plume occurs in the NH instead of the SH, as in month 6). Temperatures are once again colder in the UM but are generally colder than the differences in month 6 (i.e. month 6 differences ranged from $-24$ to 24\,K, but month 12 ranges from $-30$ to 18\,K). The UM again features a patch of warmer air at $\sim $\,60{\degree}\,N latitude up to $\sim 0.3 \sigma$ but in the opposite hemisphere to month 6. There is also a small patch of colder air at the NH pole. These temperature discrepancies are likely to be the result of no polar ice in the UM which is present in the PCM (emissivity and thermal effects of polar ice are shown by \citealp{Forget1998, Way2017}). Throughout the rest of the atmosphere, as in month 6, the UM features lower air temperatures than those in the PCM, becoming lower as the height increases up to the top of the atmosphere (Fig.~\ref{fig:temp}, month 12). Surface temperatures are also more comparable than month 9, with the only differences in temperatures $>20$\,K being at the NH pole, peaking at $>85${\degree}\,N by $\sim $\,40\,K colder in the UM.  Zonal wind differences between models are quite varied during month 12, with varied wind speed differences across the upper atmosphere in both hemispheres and a weaker and wider polar jet in the NH (Fig.~\ref{fig:zonal_winds}, month 12). The winds near the surface are quite comparable between models, but at the NH pole, the wind speeds are faster in the UM, with this difference increasing with altitude up to $\sim 0.2 \sigma$ where the difference is $\sim $\,30\,m\,s$^{-1}$. The entire NH polar jet in the UM is larger than the PCM, with the centre of the jet being spread out across more of the upper atmosphere, this is in contrast to the PCM where the NH polar jet is smaller but faster above $\sim $\,50{\degree}\,N latitude. The SH polar jet is also weaker in the UM, but this difference is much less than that of the NH (up to $-10$\,K). Meridional wind differences are also similar to month 6 in locations, but differences, where they do occur, are less than in month 3. At $\sim $\,15{\degree}\,N latitude at $\sim 0.9 \sigma$, there is now no difference between models, as opposed to during month 6, where wind speed predicted by the UM was $-2$\,m\,s$^{-1}$ slower.

Dust MMR fluctuates in both models throughout the year, becoming more abundant during month 9 (peak dust storm season, Fig.~\ref{fig:dust}) then dissipating during the colder perihelion (Fig.~\ref{fig:dust}, month 3). Both RA and RI outputs feature strong columns at the mid latitudes spanning up into the mid altitudes during dust season, with dust MMR being on average 2 orders of magnitude higher compared to that during colder months. Although there is an increase in dust MMR between seasons in both model outputs, the intensity of the change varies regionally, with poleward regions increasing less severely than equatorial regions.
\par
Despite similarities between the RA and RI scenarios, there are still some differences. Dust MMR is concentrated around the mid-latitudes for RA dust and is more dispersed with RI dust. Dust abundances are higher northward of the equator in the RI dust scenario during peak dust season (Fig.~\ref{fig:dust}, month 9), but as the dust season subsides (approaching NH spring equinox) there are higher abundances of dust MMR in the RI scenarios towards the poles (Fig.~\ref{fig:dust}, month 12). Because of this, months 6 and 12 are essential for monitoring dust-uplifting rates.

\subsection{Variable comparison to the PCM}
\label{sec:PCM}

In this section, we summarise the key differences between variables in the UM's RA simulation and the PCM output across the Martian year. We discuss zonal and meridional winds (Sect.~\ref{subsubsec:winds}), air temperature (Sect.~\ref{subsubsec:temp}) and dust (Sect.~\ref{subsec:dustdisc}). We finish the section by discussing the implications of dust differences between models and speculate as to their cause.

\hack{\newpage}

\subsubsection{Winds}
\label{subsubsec:winds}

As shown in Fig.~\ref{fig:zonal_winds}, overall patterns are similar between models, with both RA UM and PCM simulating strong eastward jets that alternate between hemispheres throughout the Martian seasons. Wind speed maxima in the PCM are generally faster than the UM, as is clearly seen in the plot for month 6, where the jets are present in both hemispheres but are $\sim $\,30\,m\,s$^{-1}$ slower in the UM. This is likely due to less atmospheric dust around $0.6\sigma $ in the UM, which in turn leads to lower temperatures; this reduces pressure gradients, causing slower winds \citep{Madeleine2011}. Despite these discrepancies, our results are encouraging, as they demonstrate the ability to model the major seasonal wind patterns with the UM.

\hack{\newpage}

As shown in Fig.~\ref{fig:meri_winds}, meridional wind patterns are similar between models but do feature some key differences. In month 3, the jet at the upper boundary of the model is situated lower in the UM, relative to the PCM output, in addition to stronger surface winds in the UM. Month 6 features the largest differences, with more distinctive jets in the UM, that are more fragmented and overall weaker in the PCM output. Months 9 and 12 are highly similar between outputs, with the largest difference being a slightly faster mid-latitude jet in the UM during month 12.

\subsubsection{Temperatures}
\label{subsubsec:temp}

As shown in Fig.~\ref{fig:temp}, the differences in temperature between the UM and PCM outputs are notable throughout the year, the highest occurring in months 6 and 12. There is a consistent difference at the poles: this is likely due to the UM not having any form of polar ice and its effect on albedo and heat transfer \citep{Forget1998, Forget1999}. These missing parameterisations will have major impacts around the near surface, as shown in Figs.~\ref{fig:temp} and \ref{fig:surftemp} at the poles, with the effect weakening with height. Month 6 and 12 differences both feature patches where the UM simulations are warmer, at  $\sim $\,60{\degree} in the hemisphere exiting winter ($\sim $\,$-60${\degree} during month 3, $\sim $\,60{\degree} during month 12). Interestingly, the differences in temperature are small in months 3 and 9 (except at the poles), despite the absence and presence of higher dust abundance, respectively. This occurs despite the difference in dust MMR maximum and minimum with month 3 being half the amount of month 9 ($6\times 10^{-6}$ and $1.2\times 10^{-5}$, respectively). This is likely due to the effects of dust uplifting and deposition (highest rate of change during months 6 and 12) having higher horizontal flux rates in the UM, leading to non-uniform differences across the atmosphere. This is also suggested by the fairly consistent distribution of differences in temperatures during months 3 and 9, where the dust is at its lowest and highest dust MMR abundances, respectively, but dust-uplifting and deposition rates are fairly homogeneous.

Across all months, the PCM output is generally warmer than the UM, with the higher temperature differences correlating with months of higher dust differences (Fig.~\ref{fig:temp}, month 9). This highlights the importance of atmospheric dust in thermal insulation in the Martian atmosphere, as temperatures nearer the upper boundary are cooler in the UM. This is likely caused by the lower amounts of suspended dust compared to the PCM, allowing more solar radiation to the surface, causing a cooling gradient as height increases above the missing dust layer \citep{Madeleine2011}.
Surface polar temperatures in the RA dust scenario are consistently different to the PCM  throughout the Martian year (Fig.~\ref{fig:surftemp}). This is most likely due to the absence of polar CO$_2$ and H$_2$O ice in the UM, which are included in the PCM. During month 9, there is still a residual ice cap in the PCM \citep{Forget1998, Forget1999}, but as there is no such cap in the UM, the radiation is incident directly on the polar soil. This will change the albedo and thermal inertia properties of these regions, leading to increased shortwave absorption in the UM than in the PCM. The subsequent effects of this potentially explain the stronger meridional circulation in the UM. As the hemispheric temperature difference is stronger, transport from the SH to the NH is amplified (as can be seen in Fig.~\ref{fig:meri_winds}).

\subsubsection{Dust content}
\label{subsec:dustdisc}

Differences in dust distributions between the UM and PCM results are the most significant of all the variables, both throughout the year and regionally within monthly outputs. Differences between the dust amount vary greatly between models, with dust season total quantities being comparable but spatially varied and the cold season retaining vastly more atmospheric dust in the PCM output than that in the RA UM scenario (Fig.~\ref{fig:dust}). Whilst the PCM output during month 3 still has less uplifted dust than its dust season quantities, there is still significantly more dust in this month than the UM by $\sim $\,1.5 orders of magnitude. This disparity changes as both models approach the dust season (Fig.~\ref{fig:dust}, month 6), however, with the dust-uplifting rate in the UM being higher than the PCM. This means that there are initially large differences between the two models, but the UM is beginning to rectify the disparity between atmospheric dust amounts as it approaches the dust season. This becomes apparent in month 9 (Fig.~\ref{fig:dust}, month 9): where the dust season is at its peak, both scenarios have much more similar amounts of dust in their outputs compared to previous months. Both outputs differ in their vertical and spatial (Figs.~\ref{fig:dust} and \ref{fig:dustsurf}) distributions, with the UM featuring a large equatorial plume that extends into the upper atmosphere, whilst the PCM features higher abundances at the SH pole. Month 9 features the first instance of a localised higher amount of near-surface dust content in the UM compared to the PCM (mainly in the Hellas Basin). Near-surface dust in the PCM, however, is distributed more evenly across the rest of the planet. This locality of dust in the UM is likely the origin of the vertical plumes present at higher altitudes (Fig.~\ref{fig:dust}). Dust is being transported southward from the equator and NH (as shown by the vectors in Fig.~\ref{fig:dustsurf}) and is redirected vertically at $\sim $\,30{\degree}\,S.
As the simulations progress towards month 12 (Fig.~\ref{fig:dust}, month 12), dust abundances in the UM reduce more quickly than the PCM, resulting in higher dust abundance in the PCM output. This is a reversal of the observed pattern as seen in month 6, where UM dust uplifting is greater than the PCM, but instead the UM dust deposition is now stronger than the PCM output. This presents an interesting dilemma in understanding how these variables are represented in simulations, as the models do not have the same parameterisations for dust uplifting, and as a result, there is a clear disparity in the amount of dust that can be uplifted between models. The usage of a ``free'' dust scheme has also been explored by \citet{Neary2018} with the Global Environmental Multiscale model (further described by \citealp{Husain2019}). They are able to characterise the Martian atmosphere according to Mars year 27 using a ``free'' dust scheme. Their work alongside this study emphasises the potential of a ``free'' dust scheme and also acts as a demonstration of model capabilities which we aim to explore further.
Although the cause of the disparity in dust distributions between those predicted by the UM and PCM is uncertain, we can speculate on potential causes. These might be caused by dust over-sensitivity to temperature in the UM. As the average atmospheric temperature decreases after the cold season, lower temperatures could potentially indirectly affect dust-uplifting or deposition rates via slower wind speeds more severely in the UM, causing uplifted dust to decrease more quickly than anticipated. There are also consistently colder air temperatures in the UM that vary in intensity throughout the Martian months, which could be amplifying this effect. Another reason might be that both models assume an infinite availability of dust reservoirs, which start uniformly distributed across the surface but are allowed to develop and congregate in areas as the model progresses. This is currently an issue in Mars modelling, as the dust deposition in ``free'' GCMs does not always match observations \citep{Montabone2020}. To rectify this, the PCM uses dust-uplifting maps to dictate where dust is being forced into the atmosphere, as described by \citet{Madeleine2011}, \citet{Spiga2013} and \citet{Montabone2020}. In our set-up, the UM does not prescribe dust uplifting in such a way but instead relies on the surface scheme to calculate dust reservoirs. Therefore, direct comparison to PCM output cannot be solely attributed to the difference in model parameterisation. Despite this, a comparison of averaged seasonal trends does show that both GCMs are able to capture annual dust storm seasons and non-dust seasons. Where dust is deposited in the model dictates source reservoirs for its subsequent uplifting. Therefore, understanding the deposition of dust and the formation of dust reservoirs (particularly after dust storms due to the amount of dust transported) is paramount to be able to reproduce the Martian dust cycle accurately \citep{Montabone2015, Montabone2017, Forget2017, Montabone2020}. If certain regions are key contributors to dust uplifting, then misrepresenting the surface wind conditions over them would have a particularly noticeable effect on the global rate of dust uplifting. The third likely cause of such differences could be dust nucleation scavenging from CO$_2$ and H$_2$O condensation being present in the PCM but absent in the UM. This affects global dust abundances where dust is being extracted from the atmosphere in the PCM during colder conditions. In the present paper, we focus only on dry simulations of the UM, so the dust abundance is not affected by scavenging from condensation, leading to potential overestimation of dust abundance during the colder cloud season. The total dust abundance during month 9 is also impacted by dust uplifting in prior months, so where pressures vary between models (Fig.~\ref{fig:surf_pres}), the ability for increased/decreased rates of dust suspension will be likely to vary between models.

Rapid vertical dust-uplifting ``rocket dust storms'', as described by \citet{Spiga2013}, play a key role in dust injection into the atmosphere. The PCM  currently factors for this, but the UM dust scheme has not previously been required to simulate such intense vertical uplifting (since it does not feature on the same scale on Earth), and therefore additional changes to the dust scheme are required in the UM. Observations and model comparison investigating this uplifting rate are essential to fine-tune the UM and verify whether the UM captures this correctly, alongside determining which developments or adjustments might be required for the UM dust scheme \citep{Madeleine2011}.

Further differences may be caused by the PCM featuring a varying inter-annual dust content between Martian years \citep{Montabone2015}, while the UM inter-annual dust content remains largely the same as that displayed in Figs.~\ref{fig:dust} and \ref{fig:dustsurf}. This was mainly mitigated by using the average scenario for the MCD (described in Sec.~\ref{sec:setup}). This remains something to consider, however, were the UM output to be used to investigate the Martian climate across multiple Martian years.

Disparities between dust MMR in the RA and PCM outputs are also affected by the absence/presence of a dust devil parameterisation. Dust devils play a large role in dust vertical transport, particularly during the NH summer \citep{Newman2002a, Kahre2006a, Kahre2017}, enabling and sustaining dust suspension above the surface (but mainly below $\sim $\,8\,km) during NH spring and summer \citep{Heavens2011, Neakrase2016, Gronoff2020, Newman2022}. While the PCM includes parameterisation for this \citep{Newman2002a}, the UM does not currently feature any explicit parameterisation for rapid vertical uplifting other than by aeolian-driven processes \citep{Marticorena1995, Woodward2011}.

Both schemes omit some dust microphysics due to the difficulty in accurately characterising them within the dust scheme, namely surface crusting and surface re-entrainment \citep{Woodward2001, Wolff2009, Madeleine2011, Woodward2022}, which likely contributes to the disparity between GCM outputs (without forcing) and observations. The differences in spatial distribution could also be caused by the differences in microphysics parameterisation between models, and therefore refining this would undoubtedly improve the representation of dust on Mars. The magnitude of this potential improvement, however, will remain uncertain until an inter-GCM comparison takes place where the initial and boundary conditions are identical. Such work would be able to identify differences which are solely due to these differences in parameterisations. Studies have been conducted on a mesoscale level in conjunction with the Mars 2020 lander \citep{Newman2022} but took place prior to recent major improvements in parameterisations in current Mars GCMs \citep{Kass2003, Madeleine2011, Spiga2013, Colaitis2013a, Navarro2014, Gonzalez-Galindo2015}. Global comparisons have been extensively applied to Earth GCMs through CMIP6 projects \footnote{Full list of projects available at \url{https://www.wcrp-climate.org/modelling-wgcm-mip-catalogue/modelling-wgcm-cmip6-endorsed-mips} (last access: 16~January~2023).} \citep{Eyring2016}. It has also been recently been conducted for exoplanets as part of the THAI project (described by \citealp{Turbet2021, Sergeev2021, Fauchez2021}), and the results have already identified new avenues for model improvement. If such a comparison was applied to Mars GCMs, it would allow for the identification of limitations of Mars modelling, potentially identifying limitations in our parameterisations, in turn allowing us to then improve comparisons to observations.

\conclusions[Discussion and conclusions]
\label{sec:conclusion}

By using multiple GCMs to simulate Mars' climate in different ways, we are able to understand areas of inaccuracy within these models \citep{Gonzalez-Galindo2010, Hinson2014, Newman2021}, as has been the case for Earth (e.g. \citealp{Eyring2016}), other solar system bodies (e.g. \citealp{Lora2019}) and  exoplanets \citep{Turbet2021, Sergeev2021, Fauchez2021}. With the UM we are able to simulate the same Martian climate with a new modelling framework. These differences are present down to the core structure of the GCMs, with the UM being a non-hydrostatic model, different parameterisations for dust calculations and a height-based vertical structure. Using a non-hydrostatic model is especially relevant to Mars, as it features periodic pressure fluctuations which affect the entire climate in multiple ways (e.g. varying wind speeds and dust-uplifting rates due to less atmospheric mass, and a summary of key differences between the models is further highlighted by \citealp{Turbet2021}). Modelling comparable climates in different GCMs is crucial for identifying differences between them, potentially like those caused by parameterisation or as a result of the different methods used to calculate variables. This work and the developments with other Mars GCMs will undoubtedly allow us to eventually expand our capabilities in areas that currently elude us, such as being able to forecast when global dust storms will occur.

The UM is capable of reproducing salient features of the large-scale circulation but lacks two key physical processes which have a considerable impact on Mars' climate: water and CO$_2$ cycle. Including parameterisations for these processes is expected to further reduce the disparities between the UM and PCM. Our goal, however, is not to make our model identical to the PCM but to offer a new modelling framework that can complement the PCM (and other GCMs) while aiming for improvements in the aspects of Mars climate where current models struggle.

Firstly, by adding water vapour and radiatively active clouds, which would affect temperatures in a variety of ways \citep{Navarro2014, Steele2017, Pal2019}, in addition to dust-uplifting rates (due to water acting as condensation nuclei for dust particles). The inclusion of these parameterisations would undoubtedly alter how the UM simulates Mars' dust and subsequently the planet's surface, especially as dust deposition changes surface properties such as thermal inertia or albedo \citep{Bonev2008, Schmidt2009, Kahre2010, Forget2017}. Although H$_2$O content is relatively low in the Martian atmosphere compared to other atmospheric compounds (even lower than Earth when accounting for the difference in atmospheric mass), it still is shown to have a large effect on Mars' climate. Radiatively active clouds can affect temperatures by up to 20\,K \citep{Madeleine2012, Navarro2014, Cooper2021}, and moisture also affects dust nucleation and deposition \citep{Walters2019}. In the present study, we use the UM with a completely dry atmosphere similar to \citet{Turbet2021}. However, as the UM has been originally developed for Earth, it already has two sophisticated cloud schemes which are routinely used for climate and weather prediction (described by \citealp{Wilson2008a, Wilson2008b}). Therefore, adding clouds to our set-up would be a matter of adaptation of an existing scheme rather than creating one from the ground up. Adding the hydrological cycle would affect dust deposition rates, especially during the colder months, when Mars' relative humidity is at its highest. Temperature profiles would be different across the atmosphere (which will cause secondary effects on winds and dust MMR) as clouds influence radiation transfer. An example of this can be seen in \citet{Navarro2014} and the scenarios in the PCM.

Secondly, our model needs to include a CO$_{2}$ cycle which substantially affects Mars' atmospheric pressure and the air--surface interaction at the poles in particular (as shown by \citealp{Forget1999, Way2017}). CO$_{2}$ condensation and sublimation lead to pressure fluctuations throughout the Martian year (Fig.~\ref{fig:pres_VL}), so they have to be accounted for by the model parameterisations to correctly reproduce horizontal pressure gradients, and thus the wind patterns \citep{Haberle2008, Read2015, Martinez2017}. Improving the atmospheric pressure characterisation will also likely improve the accuracy of dust-uplifting rates during months 6 and 12, when the surface pressure is the most different between models (Fig.~\ref{fig:surf_pres}). In some GCMs, this has been tackled by fixing the available mass of atmospheric CO$_{2}$ to match the amount for the given pressure amount, which has allowed models to characterise an incredibly complicated process, enabling an idealised representation of Mars \citep{Forget1999}. More recently, work by \citet{Way2017} has been able to alter pressure levels throughout the simulation without such ad hoc prescription. In follow-up work, we are planning to implement this or a similar parameterisation, and we expect this would improve year-round simulation as all prognostic variables shown in this paper would be affected by pressure variations, particularly during the colder months, where pressure difference between \textit{Viking} lander observations and the UM are highest. Once these processes have been added, further refinement of the albedo and surface inertia can be implemented. That is, an albedo that varies spatially across the Martian surface and is affected by CO$_2$ ice \citep{Kieffer1977, Schmidt2009}, and a thermal inertia map that also varies spatially \citep{Kieffer1977, Palluconi1981, Mellon2008}
Despite the absence of the aforementioned parameterisations, including a dust devil parameterisation and prescribed dust quantities in the UM RA, our model still produces a high-altitude dust layer using a free dust scheme. This offers a promising development in Martian climate modelling \citep{Montabone2017, Montabone2020}. While the dust quantities and their seasonality in the UM RA are not entirely similar to those in PCM (e.g. the UM RA features a single dust storm season, while Mars features two seasons in reality; \citealp{Madeleine2011, Read2015, Martinez2017}), the ability to simulate seasonal dust levels with distributions characteristic of the PCM without forcing emphasises the scientific relevance of the UM. Once more parameterisations are implemented (as mentioned in the previous two paragraphs), results may be better matched across diurnal and monthly cycles, allowing further work investigating these temporal periods. For this reason, we hope the UM will prove a vital tool in the further research of the Martian climate using GCMs.

In this paper, we have shown the first application of the UM to a modern-day Mars climate, using a dry set-up. We have demonstrated how we can adapt a highly sophisticated Earth climate model to simulate a climate on another planet. The UM demonstrates comparable wind patterns and temperature profiles to outputs from an established three-dimensional Mars GCM, the PCM. We have shown how the UM is able to simulate seasonal temperature variations and their subsequent effects on winds. We have shown how the UM can simulate uplifted dust and identified areas of disparity during colder months where the absence of a CO$_{2}$ ice and hydrological scheme likely play a role. Future work will seek to use the existing moist physics in the UM as well as to implement a CO$_{2}$ condensation scheme, allowing for the interaction of these processes with dust -- thus bringing more realism into our Mars simulations. Once these additional processes are implemented, the UM could be used to conduct simulations of specific Martian years (as done in \citealp{Montabone2015, Montabone2020}), investigate diurnal tides \citep{Hinson2004, Chapman2017, Atri2022} or provide an additional tool in the refinement of our characterisation of the Martian climate in simulations.

\hack{\newpage}

\appendix

\section{}

\begin{figure}[h!]
\hack{\hsize\textwidth}
    \includegraphics[width=14cm]{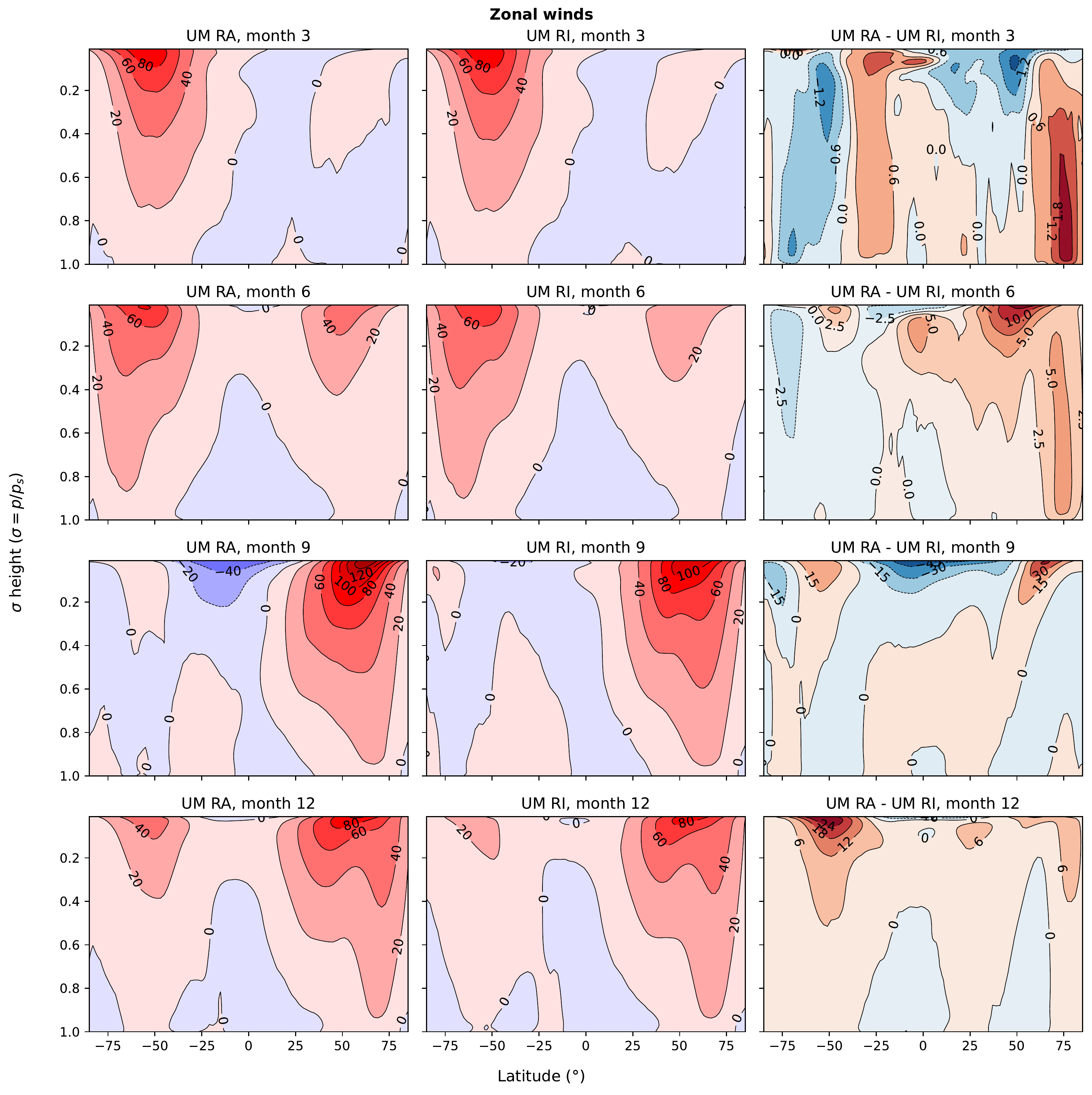}
    \caption{Zonal mean zonal winds (m\,s$^{-1}$) across four seasons within the Martian year. For each month, the time average is taken of all sols within that month. The RA dust scenario is shown on the left, the RI dust scenario in the centre and the differences between scenarios on the right. Colour scales in the left-hand and right-hand plots are matched across all months and between scenarios, with contour intervals of 20\,m\,s$^{-1}$. The contours in the difference plots are not matched due to the varying intensity of the difference between months. Positive values indicate a northward wind and negative values a southward wind.}
    \label{ap:RI_xwinds}
\end{figure}

\begin{figure}[h!]
\hack{\hsize\textwidth}
    \includegraphics[width=14cm]{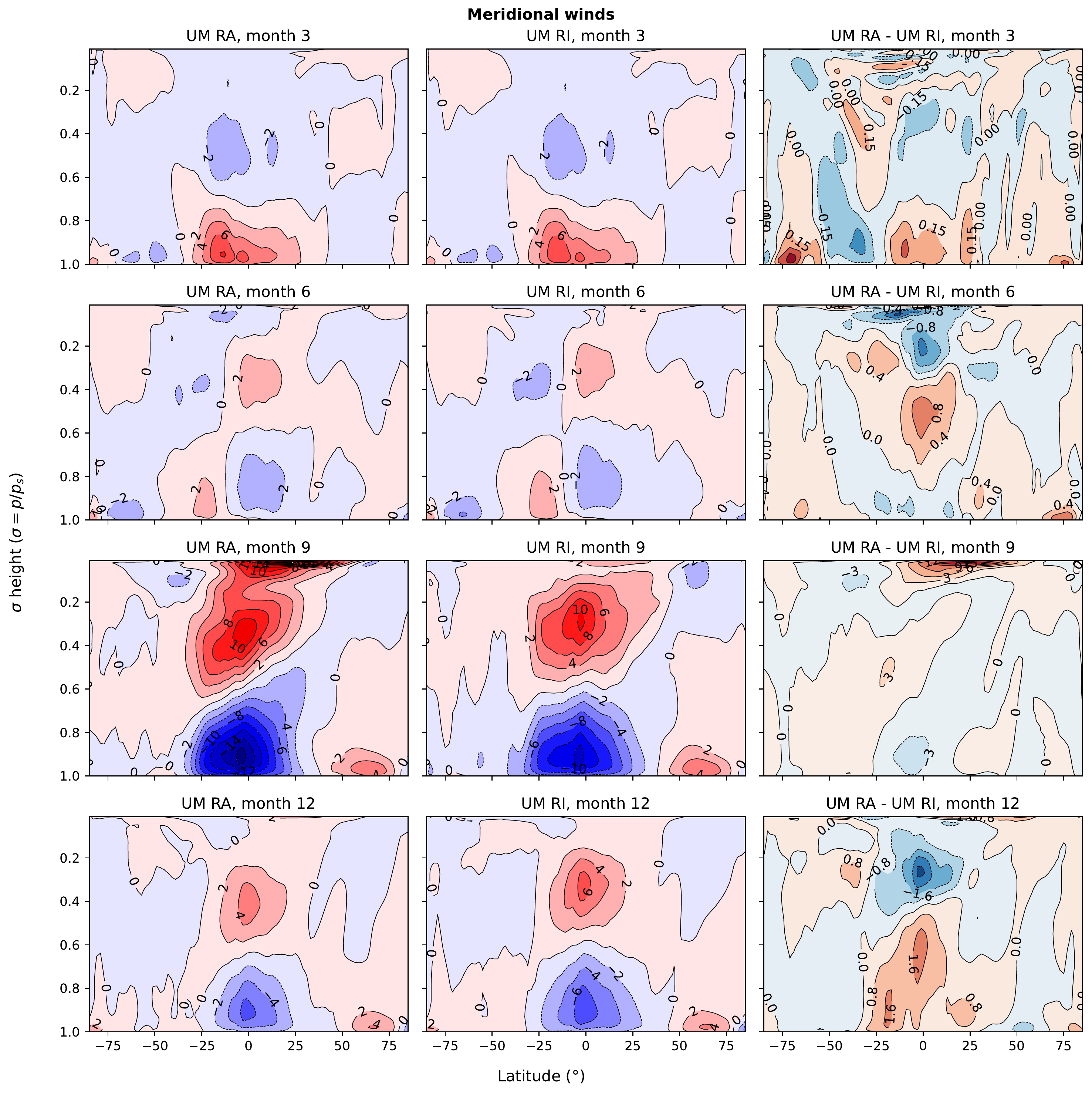}
    \caption{Zonal mean meridional winds (m\,s$^{-1}$) across four seasons within the Martian year. For each month, the time average is taken of all sols within that month. The RA dust scenario is shown on the left, the RI dust scenario in the centre and the differences between scenarios on the right. Colour scales in the left-hand and centre plots are matched across all months and between scenarios, with contour intervals of 2\,m\,s$^{-1}$. The contours in the difference plots are not matched due to the varying intensity of the difference between months. Positive values indicate an eastward wind and negative values a westward wind.}
    \label{ap:RI_ywinds}
\end{figure}

\begin{figure}[h!]
\hack{\hsize\textwidth}
    \includegraphics[width=14cm]{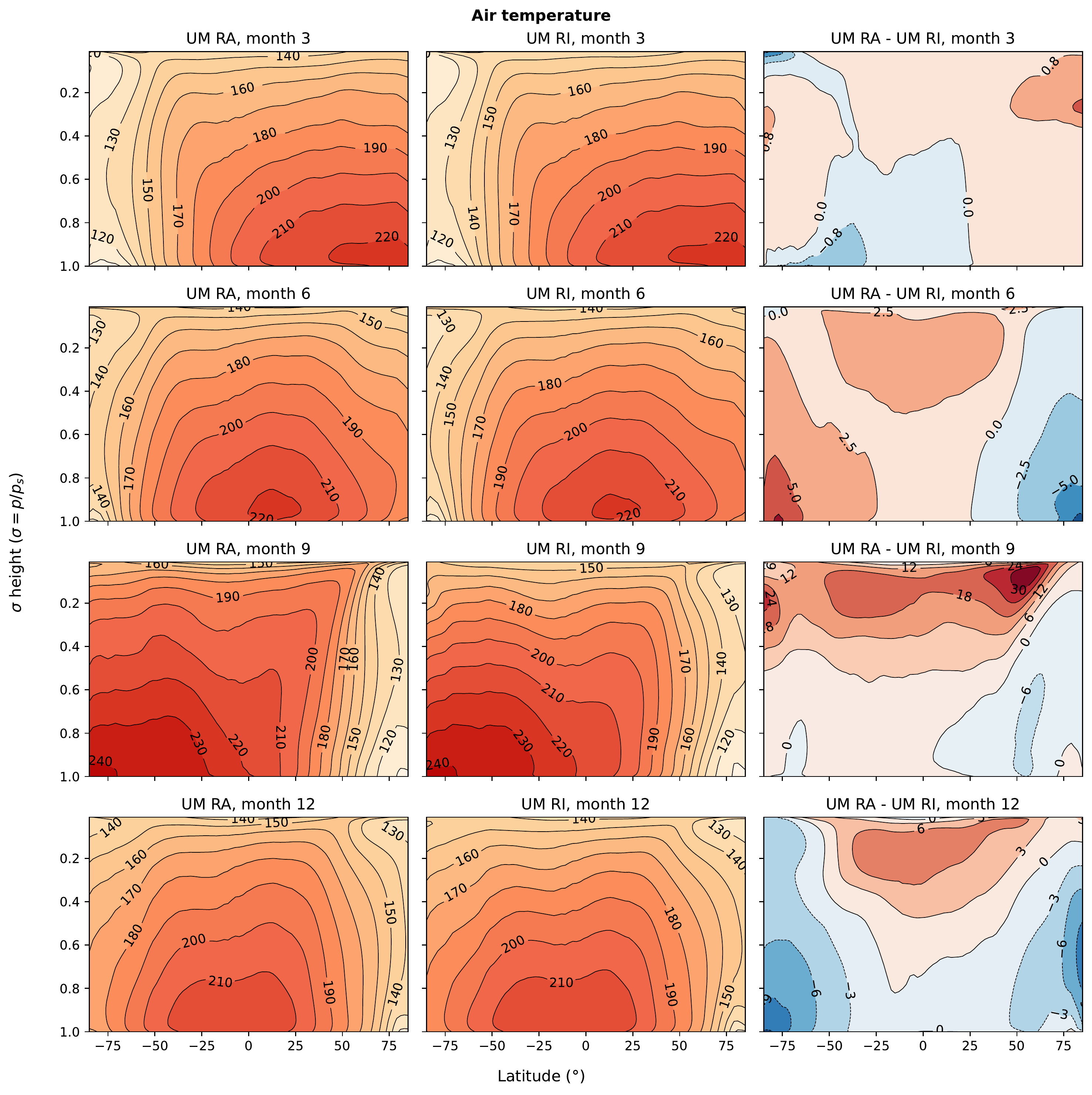}
    \caption{Zonal mean air temperature (K) across four seasons within the Martian year. For each month, the time average is taken of all sols within that month. The RA dust scenario is shown on the left, the RI dust scenario in the centre and the differences between scenarios on the right. Colour scales in the left-hand and right-hand plots are matched across all months and between scenarios, with contour intervals of 10\,K. The contours in the difference plots are not matched due to the varying intensity of the difference between months.}
    \label{ap:RI_temp}
\end{figure}

\begin{figure}[h!]
\hack{\hsize\textwidth}
    \includegraphics[width=14cm]{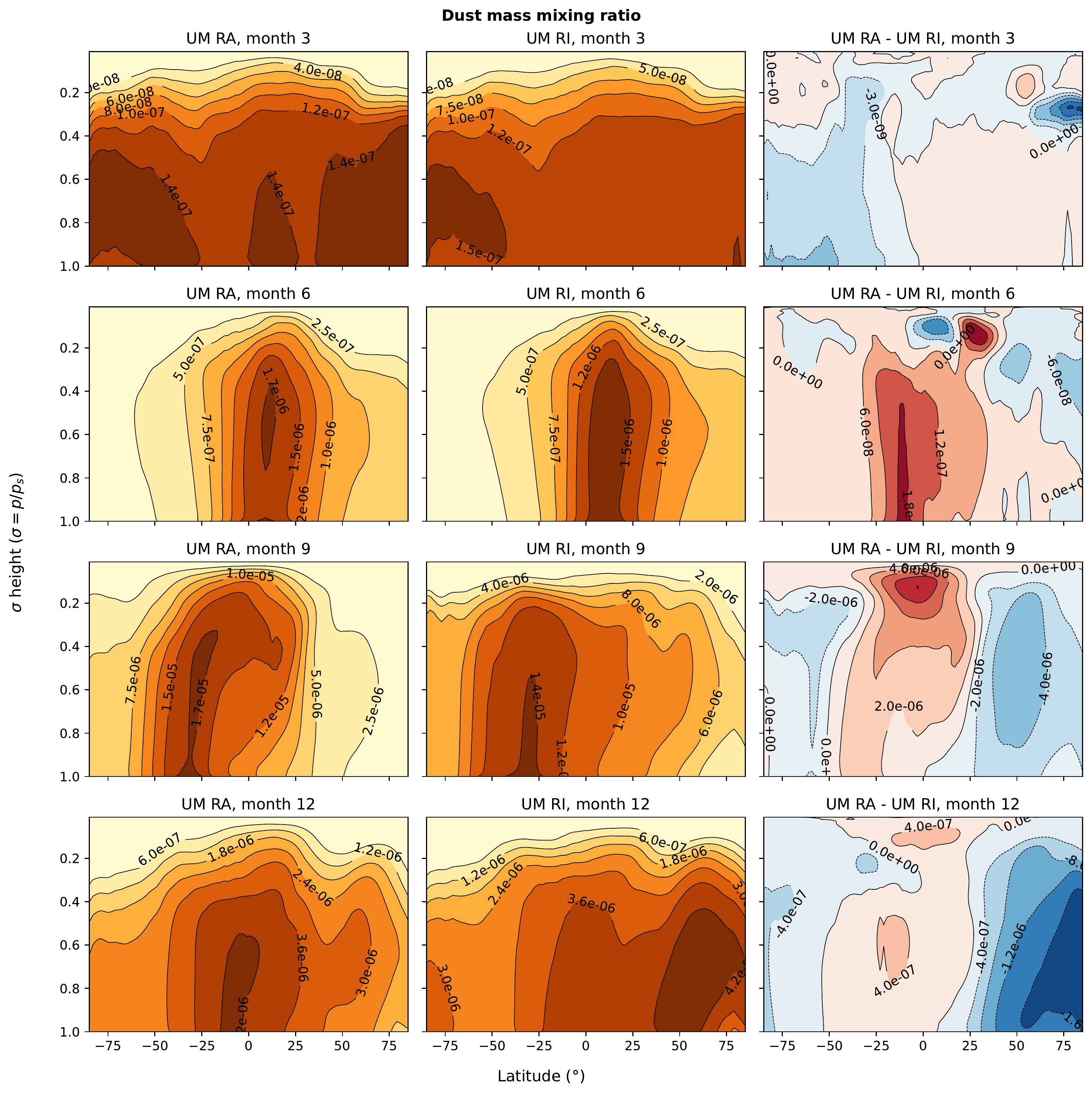}
    \caption{Zonal mean dust mass mixing ratio (kg\,kg$^{-1}$) across four seasons within the Martian year; each month is the average taken of all sols within that month (months according to Table~\ref{tab:months}). The RA dust scenario is shown on the left, the RI dust scenario in the centre and the differences between scenarios on the right. Contour lines denote the mass mixing ratio and units are in kg\,kg$^{-1}$. Note that, due to the wide range of values present between months, the colour-scale ranges differ between months and scenarios for this figure.}
    \label{ap:RI_dust}
\end{figure}

\begin{figure}[h!]
\hack{\hsize\textwidth}
    \includegraphics[width=15cm]{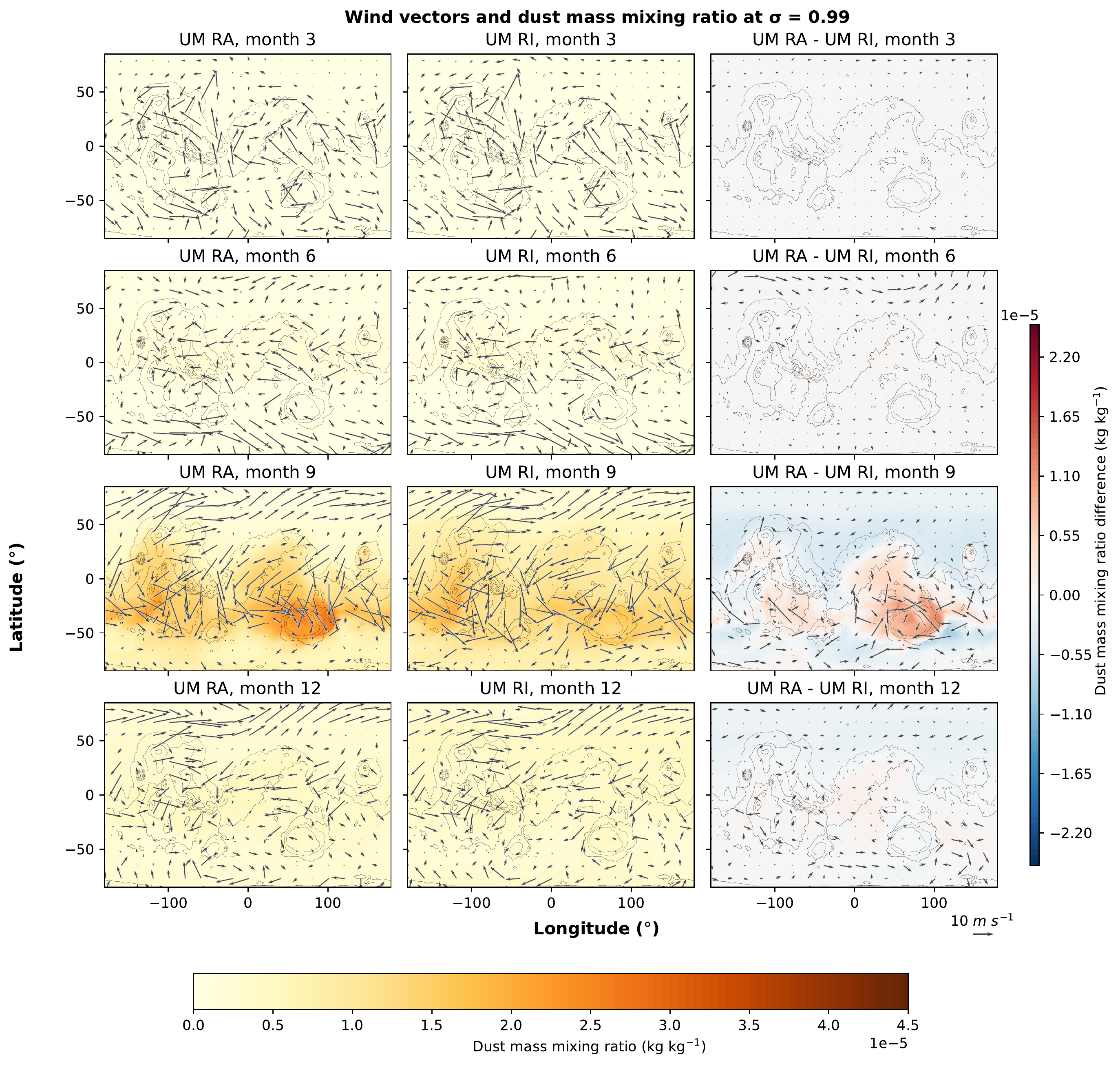}
    \caption{Dust mass mixing ratio (kg\,kg$^{-1}$) and horizontal wind speed (m\,s$^{-1}$) at $\sigma=0.99$ across four seasons within the Martian year; each month is the average taken of all sols within that month (months according to Table~\ref{tab:months}). The RA dust scenario is shown on the left, the RI dust scenario in the centre and the differences between scenarios on the right. Colour scales in the left-hand and centre plots are matched across all months and between scenarios. The contours in the difference plots are matched across months.}
    \label{ap:dustsurf_RI}
\end{figure}

\begin{figure}[h!]
\hack{\hsize\textwidth}
    \includegraphics[width=14cm]{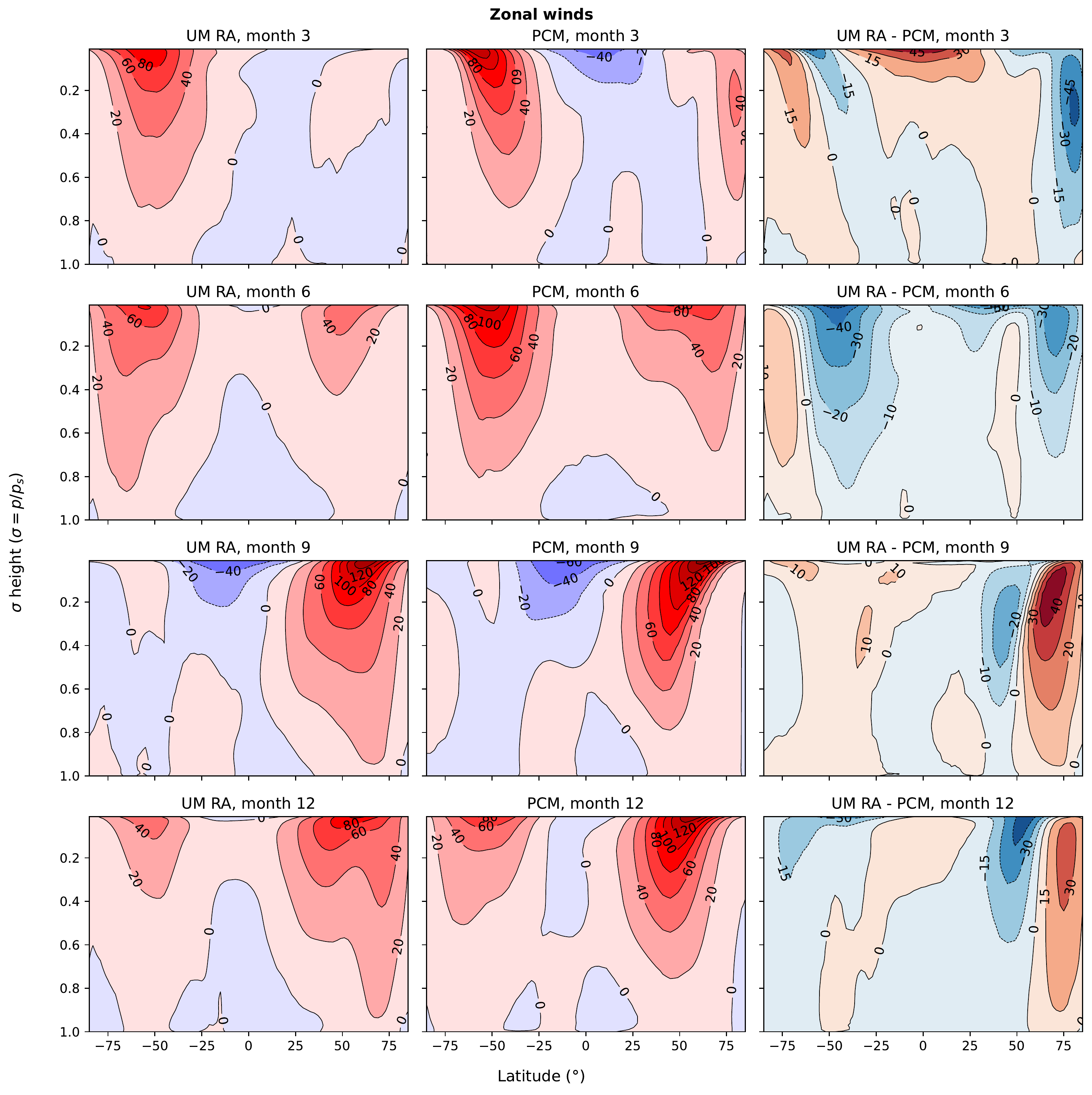}
    \caption{Zonal mean zonal winds (m\,s$^{-1}$) across four seasons within the Martian year. For each month, the time average is taken of all sols within that month. The RA dust scenario is shown on the left, the PCM output in the centre and the differences between models on the right. Colour scales in the left-hand and right-hand plots are matched across all months and between models, with contour intervals of 20\,m\,s$^{-1}$. The contours in the difference plots are not matched due to the varying intensity of the difference between months. Positive values indicate a northward wind and negative values a southward wind.}
    \label{ap:PCM_xwinds}
\end{figure}

\begin{figure}[h!]
\hack{\hsize\textwidth}
    \includegraphics[width=14cm]{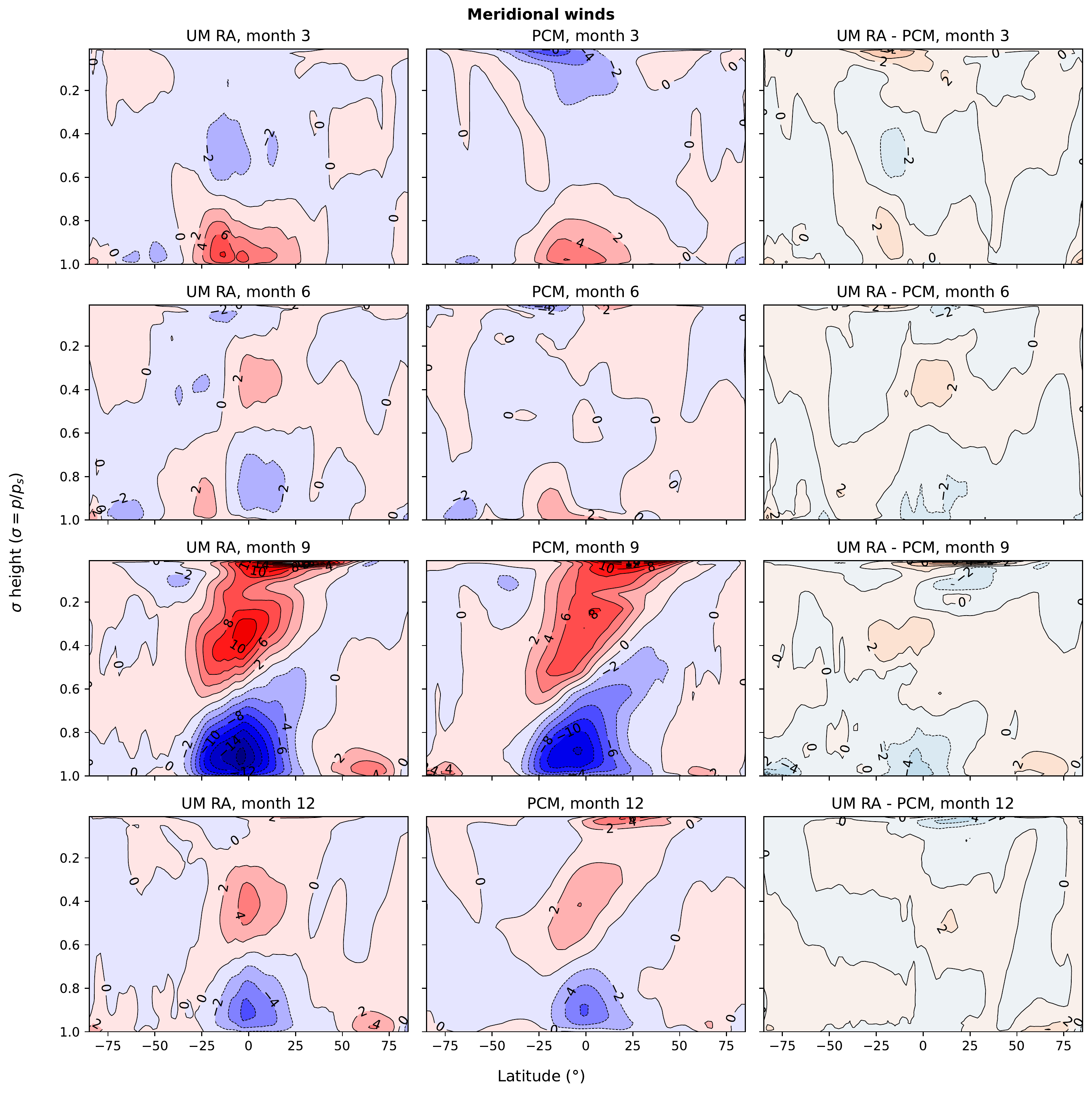}
    \caption{Zonal mean meridional winds (m\,s$^{-1}$) across four seasons within the Martian year. For each month, the time average is taken of all sols within that month. The RA dust scenario is shown on the left, the PCM output in the centre and the differences between models on the right. Colour scales in the left-hand and right-hand plots are matched across all months and between models, with contour intervals of 2\,m\,s$^{-1}$. The contours in the difference plots are not matched due to the varying intensity of the difference between months. Positive values indicate an eastward wind and negative values a westward wind.}
    \label{ap:PCM_ywinds}
\end{figure}

\begin{figure}[h!]
\hack{\hsize\textwidth}
    \includegraphics[width=14cm]{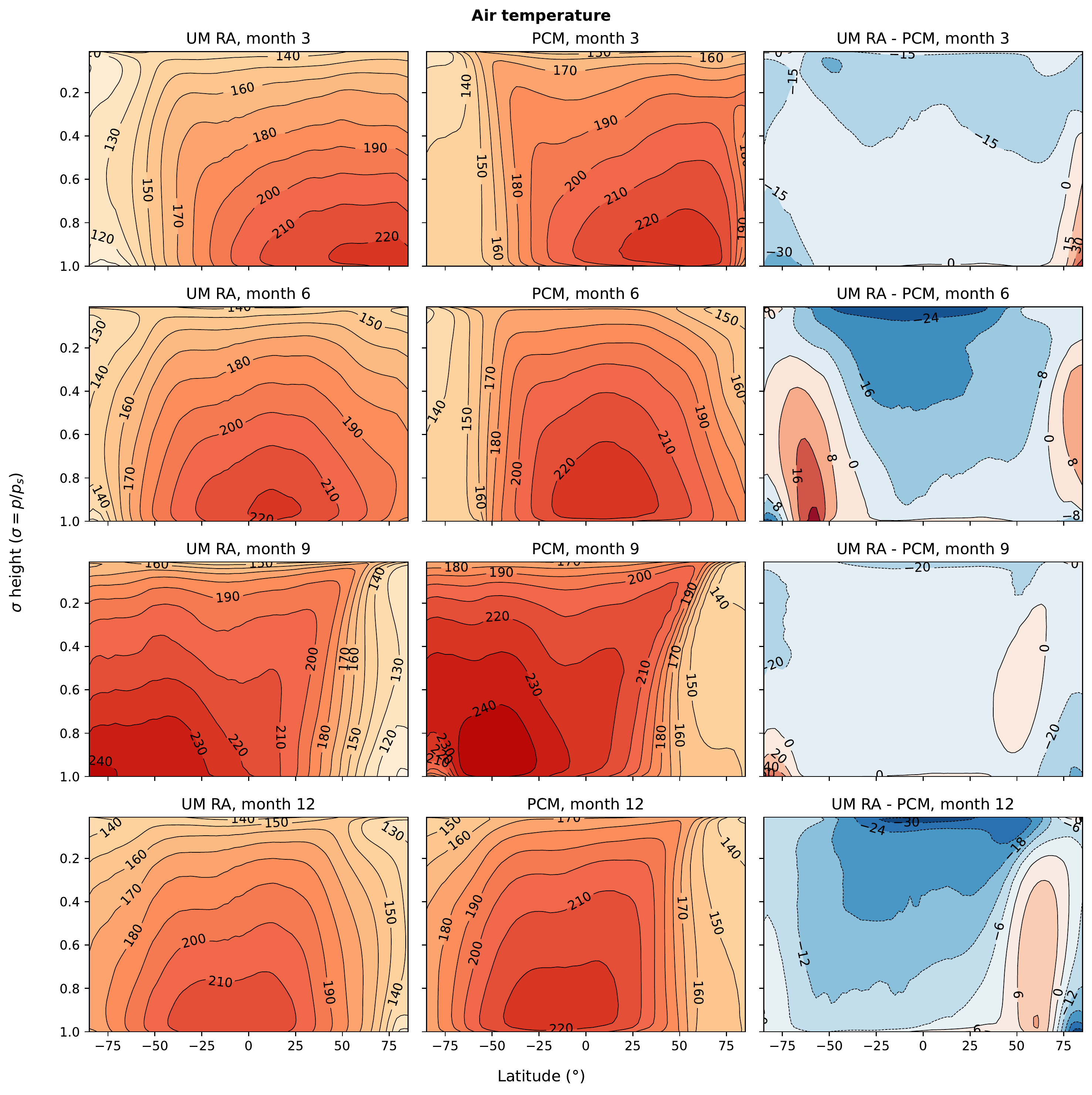}
    \caption{Zonal mean air temperature (K) across four seasons within the Martian year. For each month, the time average is taken of all sols within that month. The RA dust scenario is shown on the left, the PCM output in the centre and the differences between models on the right. Colour scales in the left-hand and right-hand plots are matched across all months and between models, with contour intervals of 10\,K. The contours in the difference plots are not matched due to the varying intensity of the difference between months.}
    \label{ap:PCM_temp}
\end{figure}

\begin{figure}[h!]
\hack{\hsize\textwidth}
    \includegraphics[width=14cm]{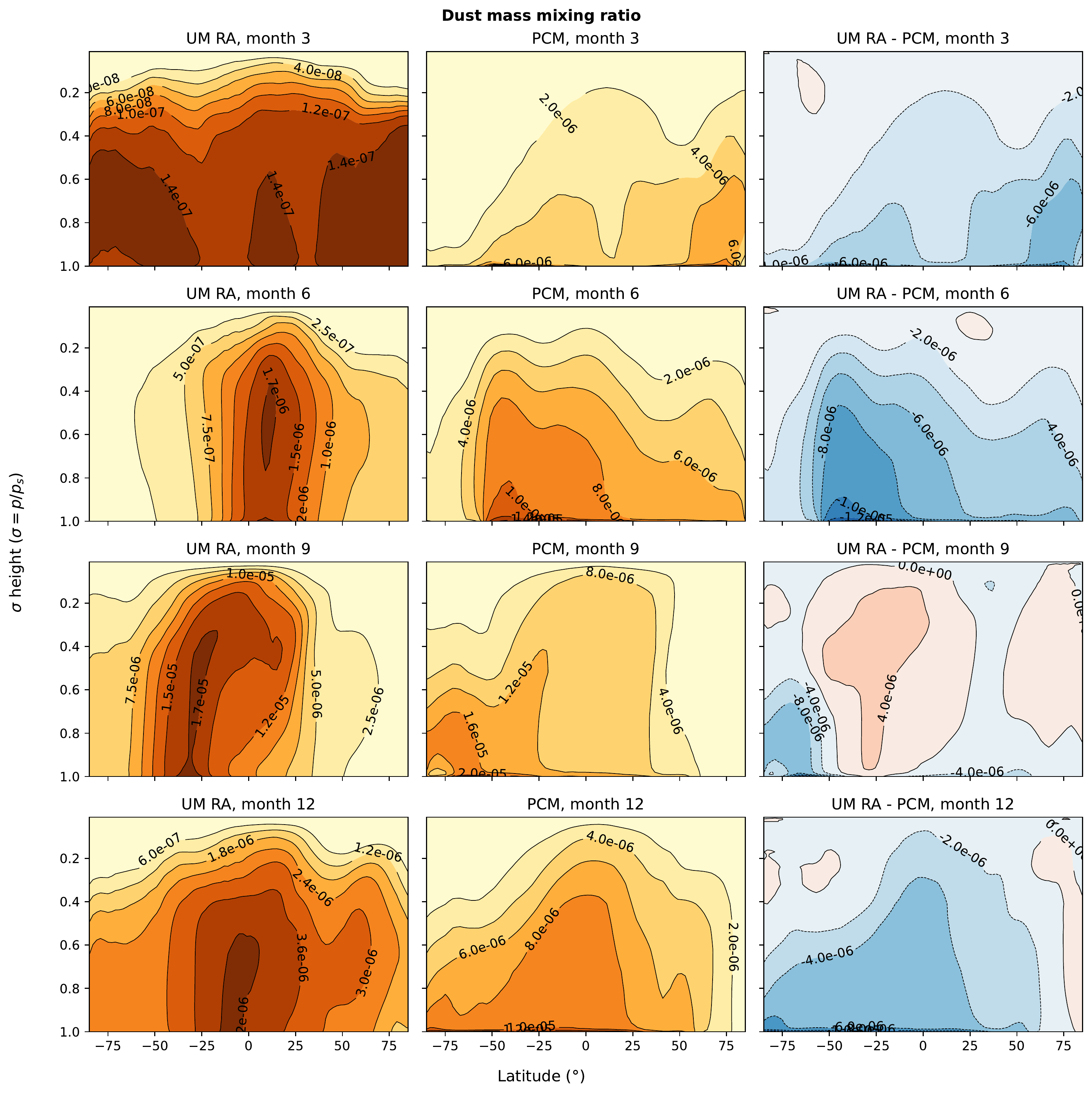}
    \caption{Zonal mean dust mass mixing ratio (kg\,kg$^{-1}$) across four seasons within the Martian year; each month is the average taken of all sols within that month (months according to Table~\ref{tab:months}). The RA dust scenario is shown on the left, the PCM output in the centre and the differences between models on the right. Contour lines denote the mass mixing ratio and units are in kg\,kg$^{-1}$. Note that, due to the wide range of values present between months, the colour scale ranges differ between months and scenarios for this figure.}
    \label{ap:PCM_dust}
\end{figure}

\begin{figure}[h!]
\hack{\hsize\textwidth}
    \includegraphics[width=15cm]{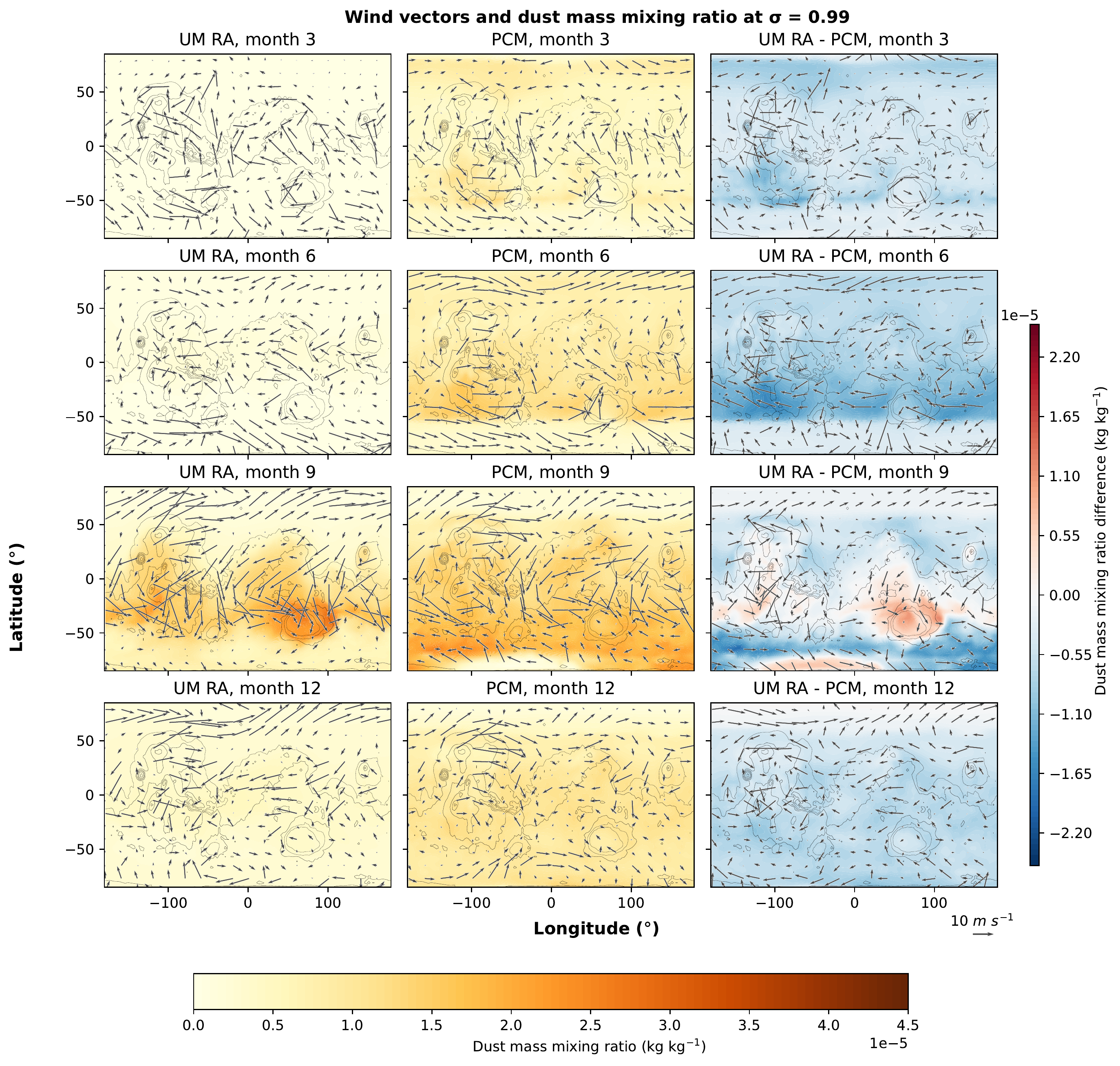}
    \caption{Dust mass mixing ratio (kg\,kg$^{-1}$) and horizontal wind speed (m\,s$^{-1}$) at $\sigma=0.99$ across four seasons within the Martian year; each month is the average taken of all sols within that month (months according to Table~\ref{tab:months}). The RA dust scenario is shown on the left, the PCM output in the centre and the differences between models on the right. Colour scales in the left-hand and centre plots are matched across all months and between models. The contours in the difference plots are matched across months.}
    \label{ap:dustsurf_PCM}
\end{figure}

In this section, we include a reference table for matching Martian months, a reference table for vertical level height  and supplementary plots for Figs.~6--9 that compare two outputs at a time (as opposed to the three in the main section of this work). In Table~\ref{tab:months}, months can be matched with the respective solar longitude and number of sols in that month. Mars features months that vary in their number of sols due to its orbital eccentricity, with fewer sols per month nearer perihelion and more sols per month closer to aphelion. Key months used in this study are months 3, 6, 9 and 12. In Table~\ref{tab:vertlevs}, vertical levels used in this configuration are given. Vertical levels are compressed/expanded depending on orography at any given point. In Figs.~\ref{ap:RI_xwinds} through \ref{ap:RI_dust} we show the RA and RI scenario outputs in the left and centre columns, respectively, with the difference between the two in the right column. In Figs.~\ref{ap:dustsurf_RI} through \ref{ap:dustsurf_PCM} we show the UM RA and PCM model outputs in the left and centre columns, respectively, with the difference between the two in the right column.

\hack{\clearpage}

\begin{table}[h!]
    \caption{Martian months, corresponding solar longitude ($L_\mathrm{s}$) and number of sols within that month. $0{\degree} L_\mathrm{s}$ corresponds to Northern Hemisphere spring equinox.}
    \label{tab:months}
    \begin{tabular}{lrrr}
    \tophline
    Month & $L_\mathrm{s}$ & Sols & Sols of month \\
    \hhline
        1 & 0--30    & 61 & 0--61\\
        2 & 30--60   & 66 & 61--127\\
        3 & 60--90   & 66 & 127--193\\
        4 & 90--120  & 65 & 193--258\\
        5 & 120--150 & 60 & 258--318\\
        6 & 150--180 & 54 & 318--372\\
        7 & 180--210 & 50 & 327--422\\
        8 & 210--240 & 46 & 422--468\\
        9 & 240--270 & 47 & 468--515\\
        10& 270--300 & 47 & 515--562\\
        11& 300--330 & 51 & 562--613\\
        12& 330--360 & 56 & 613--669\\
     \bottomhline
    \end{tabular}
\end{table}

\begin{table}[h!]
    \caption{Vertical level heights used in our Mars set-up. Values are given for a point at areoid height. This format allows for higher resolution at the surface.}
    \label{tab:vertlevs}
    \scalebox{.95}[.95]{
    \begin{tabular}{lrrrr}
    \tophline
    \multicolumn{5}{c}{$\theta$ levels -- m height} \\
    \hhline
        25.432 & 101.72	&228.864&	406.872	&635.744 \\
        915.464&	1246.056&	1627.496&	2059.8&	2542.968\\
        3076.992&	3661.872&	4297.616&	4984.216&	5721.672\\
        6509.992&	7349.176&	8239.208&	9180.112&	10\,171.864\\
        11\,214.48&	12\,307.96&	13\,452.288&	14\,647.488&	15\,893.536\\
        17\,190.456&	18\,538.224&	19\,936.856&	21\,386.344&	22\,886.696\\
        24\,387.088&	25\,888.032&	27\,391.424&	28\,901.376&	30\,425.168\\
        31\,974.128&	33\,564.536&	35\,218.528&	36\,964.968&	38\,840.408\\
        40\,889.92&	43\,168.056&	45\,739.696&	48\,680.976&	52\,080.208\\
        56\,038.736&	60\,671.856&	66\,109.72&	72\,498.248&	80\,000\\
    \hhline
    \multicolumn{5}{c}{$\rho$ levels -- m height} \\
    \hhline		
        12.712&	63.576	&165.296&	317.872&	521.312\\
        775.608&	1080.76&	1436.776&	1843.648&	2301.384\\
        2809.976&	3369.432&	3979.744&	4640.912&	5352.944\\
        6115.832	&6929.584&	7794.192&	8709.656&	9675.984\\
        10\,693.176&	11\,761.216&	12\,880.128&	14049.888&	15270.512\\
        16\,541.992&	17\,864.336&	19\,237.544&	20\,661.6&	22\,136.52\\
        23\,636.888&	25\,137.56&	26\,639.728&	28\,146.4&	29\,663.272\\
        31\,199.648&	32\,769.336&	34\,391.528&	36\,091.744&	37\,902.688\\
        39\,865.168&	42\,028.992&	44\,453.872&	47\,210.336&	50\,380.592\\
        54\,059.472&	58\,355.296&	63\,390.784&	69\,303.984&	76\,249.128\\
\bottomhline
    \end{tabular}}
\end{table}

\hack{\newpage}

\begin{table}[h!]
\caption{Dust refractive index used, as described in \citet{Balkanski2007}.}
\label{tab:dustrefractindex}
\scalebox{.80}[.80]{
    \begin{tabular}{lrrr}
    \tophline
    \multicolumn{4}{c}{Dust refractive index} \\
    \hhline
    Number & Wavelength (m) & Real part & Imaginary part \\
    \hhline
    1 & $2.00\times 10^{-7}$ & 1.520 & 0.001560 \\
    2 & $2.50\times 10^{-7}$ & 1.520 & 0.001560 \\
    3 & $3.00\times 10^{-7}$ & 1.520 & 0.001560 \\
    4 & $3.37\times 10^{-7}$ & 1.520 & 0.001550 \\
    5 & $4.00\times 10^{-7}$ & 1.520 & 0.001620 \\
    6 & $4.88\times 10^{-7}$ & 1.520 & 0.001710 \\
    7 & $5.15\times 10^{-7}$ & 1.520 & 0.001540 \\
    8 & $5.50\times 10^{-7}$ & 1.520 & 0.001470 \\
    9 & $6.33\times 10^{-7}$ & 1.520 & 0.001540 \\
    10 & $6.94\times 10^{-7}$ & 1.520 & 0.001290 \\
    11 & $8.60\times 10^{-7}$ & 1.520 & 0.000940 \\
    12 & $1.06\times 10^{-6}$ & 1.520 & 0.000669 \\
    13 & $1.30\times 10^{-6}$ & 1.510 & 0.000601 \\
    14 & $1.54\times 10^{-6}$ & 1.510 & 0.000537 \\
    15 & $1.80\times 10^{-6}$ & 1.510 & 0.000471 \\
    16 & $2.00\times 10^{-6}$ & 1.510 & 0.000473 \\
    17 & $2.25\times 10^{-6}$ & 1.510 & 0.000495 \\
    18 & $2.50\times 10^{-6}$ & 1.510 & 0.000672 \\
    19 & $2.70\times 10^{-6}$ & 1.510 & 0.001100 \\
    20 & $3.00\times 10^{-6}$ & 1.500 & 0.001780 \\
    21 & $3.20\times 10^{-6}$ & 1.500 & 0.002250 \\
    22 & $3.39\times 10^{-6}$ & 1.500 & 0.002710 \\
    23 & $3.50\times 10^{-6}$ & 1.500 & 0.002970 \\
    24 & $3.75\times 10^{-6}$ & 1.490 & 0.003610 \\
    25 & $4.00\times 10^{-6}$ & 1.490 & 0.004270 \\
    26 & $4.50\times 10^{-6}$ & 1.450 & 0.005780 \\
    27 & $5.00\times 10^{-6}$ & 1.410 & 0.007700 \\
    28 & $5.50\times 10^{-6}$ & 1.360 & 0.009540 \\
    29 & $6.00\times 10^{-6}$ & 1.290 & 0.023400 \\
    30 & $6.20\times 10^{-6}$ & 1.240 & 0.037400 \\
    31 & $6.50\times 10^{-6}$ & 1.130 & 0.166000 \\
    32 & $7.20\times 10^{-6}$ & 1.390 & 0.080500 \\
    33 & $7.90\times 10^{-6}$ & 1.080 & 0.051200 \\
    34 & $8.20\times 10^{-6}$ & 0.792 & 0.255000 \\
    35 & $8.50\times 10^{-6}$ & 1.010 & 0.504000 \\
    36 & $8.70\times 10^{-6}$ & 1.090 & 0.552000 \\
    37 & $9.00\times 10^{-6}$ & 1.300 & 0.714000 \\
    38 & $9.20\times 10^{-6}$ & 1.380 & 0.758000 \\
    39 & $9.50\times 10^{-6}$ & 2.140 & 0.843000 \\
    40 & $9.80\times 10^{-6}$ & 2.540 & 0.631000 \\
    41 & $1.00\times 10^{-5}$ & 2.480 & 0.411000 \\
    42 & $1.06\times 10^{-5}$ & 1.950 & 0.126000 \\
    43 & $1.10\times 10^{-5}$ & 1.830 & 0.143000 \\
    44 & $1.15\times 10^{-5}$ & 1.810 & 0.135000 \\
    45 & $1.25\times 10^{-5}$ & 1.630 & 0.160000 \\
    46 & $1.30\times 10^{-5}$ & 1.720 & 0.115000 \\
    47 & $1.40\times 10^{-5}$ & 1.460 & 0.165000 \\
    48 & $1.48\times 10^{-5}$ & 1.500 & 0.124000 \\
    49 & $1.50\times 10^{-5}$ & 1.470 & 0.125000 \\
    50 & $1.64\times 10^{-5}$ & 1.250 & 0.258000 \\
    51 & $1.72\times 10^{-5}$ & 1.200 & 0.413000 \\
    52 & $1.80\times 10^{-5}$ & 1.170 & 0.593000 \\
    53 & $1.85\times 10^{-5}$ & 1.190 & 0.776000 \\
    54 & $2.00\times 10^{-5}$ & 1.420 & 0.950000 \\
    55 & $2.13\times 10^{-5}$ & 1.670 & 1.490000 \\
    56 & $2.25\times 10^{-5}$ & 2.840 & 0.874000 \\
    57 & $2.50\times 10^{-5}$ & 1.920 & 0.652000 \\
    58 & $2.79\times 10^{-5}$ & 2.070 & 0.393000 \\
    59 & $3.00\times 10^{-5}$ & 1.850 & 0.592000 \\
    60 & $3.50\times 10^{-5}$ & 1.670 & 0.538000 \\
    61 & $4.00\times 10^{-5}$ & 1.630 & 0.555000 \\
    62 & $1.00\times 10^{-2}$ & 1.630 & 0.554000 \\
    \bottomhline
    \end{tabular}}
\end{table}

\hack{\clearpage}

\hack{\clearpage}

\hack{\clearpage}

\hack{\clearpage}

\hack{\clearpage}

\hack{\clearpage}

\hack{\clearpage}

\hack{\clearpage}

\hack{\clearpage}

\hack{\clearpage}

\codeavailability{Scripts to process and visualise the post-processed UM data used in this study, alongside package requirements and tutorials, are available as a Zenodo dataset: \url{https://doi.org/10.5281/zenodo.6974260} \citep{McCulloch2022a}. If you do use those data, then please cite this paper and add the following statement: ``UM data have been obtained from \url{https://doi.org/10.5281/zenodo.6974260}''.

\hack{\newpage}
\hack{\vspace*{17.2cm}}

Due to intellectual property right restrictions, we cannot provide either the source code or documentation papers for the UM or JULES. The Met Office Unified Model is available for use under licence. A number of research organisations and national meteorological services use the UM in collaboration with the Met Office to undertake basic atmospheric process research, produce forecasts, develop the UM code, and build and evaluate Earth system models. For further information on how to apply for a licence, see \url{https://www.metoffice.gov.uk/research/approach/modelling-systems/unified-model} (last access: 18 April 2022). Obtaining JULES: JULES is available under licence free of charge. For further information on how to gain permission to use JULES for research purposes, see \url{http://jules-lsm.github.io/access_req/JULES_access.html} (last access: 3 April 2022). UM--JULES simulations are compiled and run in suites developed using the Rose suite engine (\url{http://metomi.github.io/rose/doc/html/index.html}, last access: 16~January~2023) and scheduled using the cylc workflow engine (\url{https://cylc.github.io/}, \citealp{Oliver2019}). Both Rose and cylc are available under v3 of the GNU General Public License (GPL). In this framework, the suite contains the information required to extract and build the code as well as configure and run the simulations. Each suite is labelled with a unique identifier and is held in the same revision-controlled repository service in which we hold and develop the model code. This means that these suites are available to any licensed user of both the UM and JULES.}

\dataavailability{A post-processed dataset is provided in \citet{McCulloch2022a} (\doi{10.5281/zenodo.6974260}). This dataset contains the zonally meaned outputs from the UM RA and RI scenarios. For PCM data, please contact the MCD team (\url{http://www-mars.lmd.jussieu.fr/mars/info_web/index.html}, last
access: 16~January~23).}

\authorcontribution{DM led the writing and suite development with supervision from DES, NM, and MB. JM, BD and IB provided assistance in tuning the model and provided thorough descriptions on how they work. KK provided technical support in IT infrastructure to access the model code and Monsoon2 system. The paper was reviewed and contributed to by all the co-authors.}

\competinginterests{The contact author has declared that none of the authors has any competing interests.}

\begin{acknowledgements}
We thank two anonymous reviewers for their comments that helped improve this paper.
We would like to thank the wider Exeter Exoplanet Theory Group for their feedback and support in UM development, as well as Patrick McGuire from the University of Reading for acting as a springboard for ideas. The authors also acknowledge the MCD data team: François~Forget, Aymeric~Spiga, Ehouarn~Millour, for providing freely accessible Mars climate data at \url{http://www-mars.lmd.jussieu.fr/mars/info_web/index.html} (last access: 6~March~2023). Material produced using Met Office Software. We acknowledge use of the Monsoon2 system, a collaborative facility supplied under the Joint Weather and Climate Research Programme, a strategic partnership between the Met Office and the Natural Environment Research Council. This work was partly supported
by a Science and Technology Facilities Council Consolidated
Grant (ST/R000395/1), a Leverhulme Trust research project
grant (RPG-2020-82) and a UKRI Future Leaders Fellowship (grant
no. MR/T040866/1). For the purpose of open access, the authors have applied a Creative Commons Attribution (CC BY) licence to any Author Accepted Manuscript version arising.
\end{acknowledgements}

\financialsupport{This work was partly supported by a Science and Technology Facilities Council Consolidated Grant (grant no. ST/R000395/1), a Leverhulme Trust research project grant (grant no. RPG-2020-82) and a UKRI Future Leaders Fellowship (grant no. MR/T040866/1).}

\reviewstatement{This paper was edited by Jinkyu Hong and reviewed by two anonymous referees.}

\end{document}